\documentclass[prd,aps,showpacs,twocolumn,superscriptaddress,floatfix,nofootinbib]{revtex4-1}
\usepackage[utf8]{inputenc}
\usepackage{bm}
\usepackage{bbold}
\usepackage{mathtools}
\usepackage[toc,page]{appendix}
\usepackage{dsfont}
\usepackage{combelow}
\usepackage[colorlinks=true,linkcolor=blue,citecolor=blue, urlcolor=blue]{hyperref} 
\usepackage[a4paper,top=2.4cm,bottom=2.4cm,right=1.0cm,bindingoffset=-1.2cm]{geometry}
\usepackage{tikz}
\usepackage{graphicx}
\usepackage{subfigure}
\usepackage{scalerel,amssymb}
\DeclareMathAlphabet{\pazocal}{OMS}{zplm}{m}{n}
\usepackage{ulem}

\newcommand{\trento}{\textsc{trento}}
\renewcommand{\d}{\mathrm{d}}
\newcommand{\epn}{\frac{\d E_\perp^{(0)}}{\d \eta}}
\newcommand{\epni}{{\left.\d E_\perp^0\right/\d \eta}}

\newcommand{\xT}{{\mathbf{x}_\perp}}
\newcommand{\pT}{\mathbf{p_\perp}}

\newcommand{\W}{\mathrm{W}}

\begin{document}
\title{Collective dynamics in heavy and light-ion collisions — II) Determining the origin of collective behavior in high-energy collisions}
\author{Victor E. Ambru\cb{s}}
\affiliation{Department of Physics, West University of Timi\cb{s}oara, \\
Bd.~Vasile P\^arvan 4, Timi\cb{s}oara 300223, Romania}
\author{S.~Schlichting}
\affiliation{Fakultät für Physik, Universität Bielefeld, D-33615 Bielefeld, Germany}
\author{C.~Werthmann}
\email{clemens.werthmann@ugent.be}
\affiliation{Fakultät für Physik, Universität Bielefeld, D-33615 Bielefeld, Germany}
\affiliation{Incubator of Scientific Excellence-Centre for Simulations of Superdense Fluids, University of Wrocław, pl. Maxa Borna 9, 50-204 Wrocław, Poland}
\affiliation{Department of Physics and Astronomy, Ghent University, 9000 Ghent, Belgium}
\date{\today}

\begin{abstract}
Exploiting the first measurements of the same ion species in OO collisions at RHIC and LHC, we propose an observable to distinguish whether collective behavior builds up through a hydrodynamic expansion of a strongly interacting quark-gluon plasma or few final state rescatterings. Our procedure allows one to disentangle the effects of the initial state geometry and the dynamical response mechanism on anisotropic flow. We validate its ability to discriminate between systems with different interaction rates using results from event-by-event simulations in kinetic theory.

\end{abstract}

\pacs{}
\maketitle

\section{Introduction} 
Experimental observations of collective behavior in hadronic collisions as small as pA~\cite{ALICE:2014dwt} or pp~\cite{ATLAS:2017hap,CMS:2017kcs}, have generated an intense debate on different possible origins of this phenomenon and prompted various theoretical investigations to explain the underlying mechanism
(see e.g. Ref.~\cite{Grosse-Oetringhaus:2024bwr} for a recent review). Collective flow is usually quantified by the flow harmonics $v_n$, which are defined as the Fourier coefficients of the final state one particle distribution in azimuthal angle with respect to the global symmetry plane of the event. It has been established that these flow harmonics are dynamically generated as a final state response to the initial state geometry. In particular, up to higher order corrections, flow harmonics have been observed to be in almost perfect proportionality with the initial state eccentricities measuring anisotropies with n-fold rotational symmetry in the system's position space distribution.  
\begin{align}
    v_{n} \simeq \kappa_{n,n} \epsilon_{n}
\end{align}
Since there are two factors in this equation, it is generally not possible to distinguish the case of large $\kappa$ and small $\epsilon$ from the opposite case, especially in small systems, where the initial state geometry is poorly constrained. 
Due to these uncertainties, different physics mechanisms can explain the observed flow harmonics $v_n$; among those are hydrodynamic models assuming a high rate of final state interactions and kinetic escape assuming a very low interaction rate, which can both describe the experimental data in small systems. Stated differently, it is clear that poorly constrained models of the initial state in small systems can mask inaccuracies in the modeling of the dynamical response of the system, to yield agreement with experimental measurements.

Evidently, it would be highly desirable to either obtain a more accurate theoretical description of the initial state in small systems or eliminate the uncertainties of the initial state geometry in extractions of the flow response from experimental data. Here, we propose an observable that takes advantage of collisions of the same ion species at different center of mass energies to achieve such a comparison. The interpretation of this observable is informed by our studies of the applicability of hydrodynamics in hadronic collisions. These are discussed in previous works~\cite{Ambrus:2022koq,Ambrus:2022qya}, as well as in our companion paper~\cite{Ambrus:2024hks}, which also provides further details on the setup and rationale of this work.

In this paper, we briefly introduce the setup and general idea behind the construction of the observables in Sec.~\ref{sec:separating} and validate its discriminative power via simulations in kinetic theory in Sec.~\ref{sec:validaiton}. After introducing the simulation setup and discussing the necessary prerequisites on the flow results, we perform a rigorous derivation of the observable in Sec.~\ref{sec:deriving_observables} and then compare simulation results to theory expectations in Sec.~\ref{sec:simulation_results}. In Sec.~\ref{sec:non-conformal}, we test how the observable performs in the presence of nonconformal effects and then we conclude with Sec.~\ref{sec:conclusion}. In App.~\ref{app:fwork_corr} we derive an improved theory expectation for the observable and in App.~\ref{app:hydro_curves} we test how it performs for conformal hydrodynamic simulations. The raw data for all plots presented in this work are publicly available~\cite{werthmann_2025_14849764}.

\section{Separating response and geometry}\label{sec:separating}
The key observation that motivates this work is the fact that the buildup of flow in the dense versus dilute regime can be distinguished by the qualitatively different dependence on the interaction strength. This is convenient to characterize via the dimensionless opacity measure
\begin{align}
    \hat{\gamma}=\frac{1}{5 \eta/ s} 
 \left(\frac{R}{\pi a} \frac{\d E_\perp^0}{\d\eta}\right)^{1/4}\;,\label{eq:ghat}
\end{align}
collecting parametric dependencies on the specific shear viscosity $\eta/s$, the initial transverse energy per unit rapidity $\epni$ and the initial rms transverse radius $R$ of the energy density profile.

The qualitative behaviour of the buildup of flow can be argued on very general grounds. In weakly interacting systems, the interaction can be treated as a linear perturbation to the noninteracting case and thus
\begin{align}
    \kappa(\hat{\gamma} \ll 1) \simeq \kappa'_{0} \hat{\gamma} \;  \mathrm{(linear)}.
\end{align}
When increasing the interaction strength of the system, the response will saturate and eventually converge towards the ideal hydrodynamic limit
\begin{align}
    \kappa(\hat{\gamma} \gg 1) \simeq \kappa_{\infty}  \; \mathrm{(const.)}.
\end{align}
This dependence on the opacity has effects also on the flow cumulants as a function of the centrality, as more central events have a higher interaction strength. Thus, together with the varying geometry, the centrality dependence of flow cumulants emerges from a combination of two entangled effects.

While these are indirect effects, our central idea is to investigate the same system at two different opacities by comparing RHIC and LHC results, in order to disentangle the effects of the initial state geometry $\epsilon_{n}$ from those of the final state response mechanism, encoded in $\kappa_{n,n}$. Specifically, we will demonstrate that ratios of flow cumulants are mostly determined by the geometry and thus insensitive to response, whereas the ratio of logarithmic differences in elliptic flow and transverse energy can be used to
directly assess the opacity dependence of the flow response, and thereby 
probe whether the system behaves close to an ideal fluid or not.

\subsection{Simulation setup}~\label{sec:validaiton}
Clearly, the above arguments are rather general in nature and we expect them to hold independent of the detailed dynamical description of the system. However, in order to validate this procedure and study the discriminative power of the proposed observables, we  will compute them within a specific incarnation of kinetic theory, which -- as recently demonstrated~\cite{Ambrus:2022koq} -- can describe the onset of collective flow in the full range of opacities, smoothly interpolating between both ends of the spectrum from noninteracting systems to ideal hydrodynamics. We follow this setup and employ the Boltzmann equation in conformal relaxation time approximation (RTA):
\begin{align}
    p^{\mu}\partial_{\mu} f(x,p)= -\frac{u^{\mu}(x)p_{\mu}}{\tau_R(x)} \left[f(x,p)-f_{\rm eq}(x,p)\right]\;.
\end{align}
We focus on flow observables that can directly be obtained from the energy-momentum tensor, namely the cumulants
\begin{align}
    c_{\varepsilon_p}\{2\}=\langle |\varepsilon_p|^2 \rangle\;, \qquad c_{\varepsilon_p}\{4\}= \langle  |\varepsilon_p|^4 \rangle - 2\langle |\varepsilon_p|^2 \rangle \label{eq:cumulant}
\end{align}
of the event-by-event momentum space anisotropy\footnote{Note that the final $\varepsilon_{p}$ can be obtained experimentally from an observed particle spectrum $\frac{\d N}{\d^2\pT \d y}$ as $\varepsilon_p=\left(\int_\pT \sqrt{\pT^2+m^2}~\frac{\d N}{\d^2\pT \d y} \right)^{-1}\left(\int_\pT \sqrt{\pT^2+m^2} e^{2i\phi}~\frac{\d N}{\d^2\pT \d y}\right)$ , although we believe that $v_{2}$ will give similar results. }
\begin{align}
\varepsilon_{p}=\frac{\int_\xT T^{xx}-T^{yy}+2iT^{xy}}{\int_\xT T^{xx}+T^{yy}}\;.\label{eq:epsp}
\end{align}
Compared to the usual $v_n$s, the clear advantage of these observables is that they can be computed without invoking a particular hadronization procedure. Since they measure the flow of energy, which is a conserved quantity, we can anticipate that they avoid uncertainties associated with the hadronization process.

The event-by-event initial states were generated in \trento ~using a table of 6000 pre-generated nucleon configurations of the $^{16}\mathrm{O}$ nuclei. These configurations were obtained using quantum Monte Carlo calculations based on two- and three-body potentials that were derived in chiral effective field theory~\cite{Lim:2018huo}\footnote{The list of  $^{16}\mathrm{O}$ configurations is publicly available~\cite{TGlauberMC}.}. They have already been used in other works on simulations of OO collisions~\cite{Lim:2018huo,Rybczynski:2019adt,Nijs:2021clz} and we followed Ref.~\cite{Nijs:2021clz} in setting the \trento ~parameters. Most of the results presented here focus on OO collisions, but we also present some results for PbPb and AuAu. The initial state setup for these simulations was similar. For details, we refer to our companion paper~\cite{Ambrus:2024hks}. The numerical setup for computing the time evolution is identical to the one presented in Ref.~\cite{Ambrus:2021fej}.

\begin{figure}
    \centering
    \includegraphics[width=.49\textwidth]{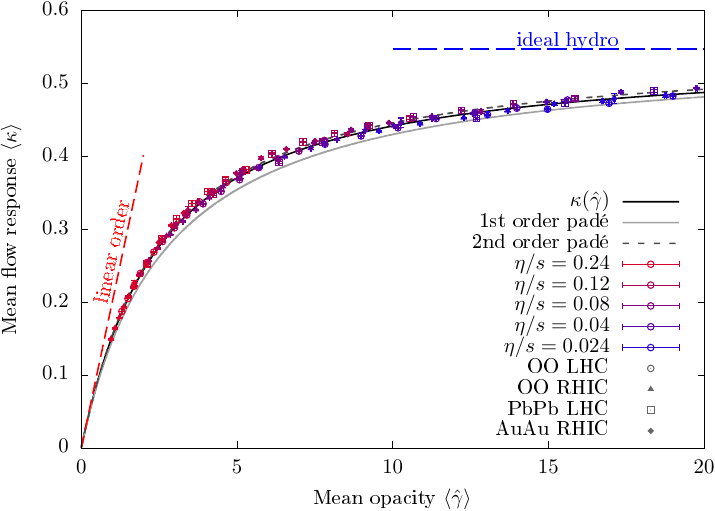}
    \caption{Mean values of the response $\kappa=\varepsilon_p/\epsilon_2$ as a function of the mean opacity from our kinetic theory simulation results for centrality classes of size 10\% of OO collisions at LHC (open circles) and RHIC (filled triangles) as well as PbPb collisions (open squares) and AuAu collisions (filled diamonds) for various values of $\eta/s$ in color gradient from highest (red) to lowest (blue), compared to a constant rescaling of the $\kappa(\hat{\gamma})$-curve from our previous work (black) as well as a first order (light gray, solid) and second order (dark gray, dashed) Padé fit to the data with the limiting cases of opacity linearized buildup for dilute systems (red dashed) and saturated response for ideal hydrodynamics (blue dashed). Error bars show the standard error of the mean.}
    \label{fig:kappa_vs_gamma}
\end{figure}

All events were simulated at two collisional energies (RHIC: $200\;$GeV, LHC: $7\;$TeV) and at five different values of the shear viscosity,  $4\pi\eta/s=0.3,0.5,1.0,1.5,3.0$, as well as in ideal hydrodynamics and an opacity-linearized time evolution setup. For a more detailed description of the setup, including also the limiting cases of opacity linearized and ideal hydrodynamic results, we refer to the companion paper~\cite{Ambrus:2024hks}.

\subsection{Universal collective flow response}
Before testing our observables, we first establish that there is indeed a universal opacity dependent elliptic flow response curve $\kappa(\hat{\gamma})$. This will also be instrumental in comparing simulation results for the observables to theory expectations derived from this curve. To do this, we assess the mean values of the flow response coefficient $\kappa=\varepsilon_p/\epsilon_2$ as a function of the mean opacity $\hat{\gamma}$ for the centrality classes of our simulation results. The result is illustrated in Fig.~\ref{fig:kappa_vs_gamma}. Different symbol types show the different collision systems and different colors show simulation results at different values of $\eta/s$.

As can be seen in Fig.~\ref{fig:kappa_vs_gamma}, all simulation results clearly line up along a common curve, indicating a remarkable degree of universality of the flow response across all considered systems. We want to extract this universal curve $\kappa(\hat{\gamma})$ by fitting Padé approximations to the data points. These Padé approximations were obtained as follows. We refer to the curves as approximations instead of fits, because the first order one is technically not a fit to the plotted data. It was constructed as an interpolation of the asymptotic behaviour of the flow response coefficient in the linear order case at small opacities ($\kappa_{\rm LO}=0.201$) and the ideal hydrodynamic case at large opacities ($\kappa_{\rm id}=0.547$). This fixes it to be of the following form.
\begin{align}
    \kappa_{1,1}(\hat{\gamma})=\frac{\kappa_{\rm LO}\hat{\gamma}}{1+\frac{\kappa_{\rm LO}}{\kappa_{\rm id}}\hat{\gamma}}
\end{align}
The second order Padé has more parameters, some of which are not fixed by the asymptotic behaviour. There is some freedom in how to choose them. We picked the following functional form. 
\begin{align}
    \kappa_{2,2}(\hat{\gamma})=\frac{\kappa_{\rm LO}\hat{\gamma}+a\kappa_{\rm id}\hat{\gamma}^2}{1+b\hat{\gamma}+a\hat{\gamma}^2}
\end{align}
The values of the additional parameters were obtained by a fit to the data to be $a=0.217(31)$ and $b=0.845(73)$.

In Fig.~\ref{fig:kappa_vs_gamma}, the black curve represents $\kappa(\hat{\gamma})$ as it was obtained from an interpolation of the results for a fixed average event from our previous study~\cite{Ambrus:2022koq}, scaled by a factor of $0.93$. The light gray and dark gray dashed lines show first and second order Padé approximations of the data. The red and blue dashed lines show the opacity linearized and ideal hydrodynamic limits.

The first order Padé underestimates the flow response curve, but the other two lines are reasonable approximations thereof, hence we will use the second order Padé to describe $\kappa(\hat{\gamma})$ in the following. When comparing to the curve obtained from the results of our previous study, the scaling factor $0.93$ might seem arbitrary. The initial condition we used there was an average of event-by-event profiles obtained from a different, saturation based initial state model. It is unclear whether the stronger flow response to that profile is due to the extraordinarily smooth profile or to the different initial state model. However, the plot shows that the factor is the only difference in the curve, while the qualitative behaviour is almost exactly the same. This means that both curves yield almost exactly the same result for ratios of flow responses in different systems.

\begin{figure*}
    \centering
    \includegraphics[width=.49\textwidth]{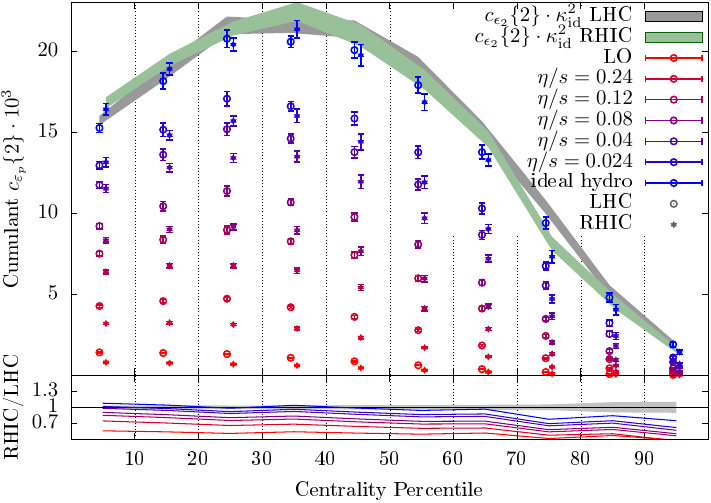}
    \includegraphics[width=.49\textwidth]{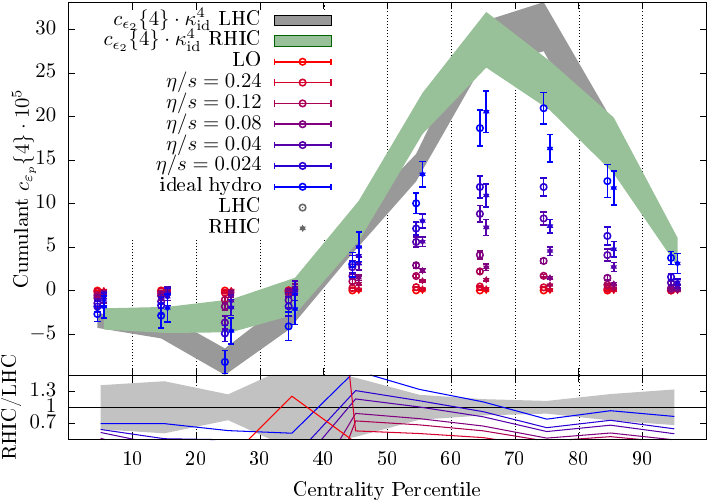}
    \caption{
    The second order (left) and fourth order (right) cumulants of elliptic flow $\varepsilon_p$ (crosses) from simulations of OO collisions at LHC (empty circles) and RHIC (filled stars) collisions at different shear viscosity (different colors) as a function of centrality. The bands show the corresponding initial state ellipticity cumulants multiplied by a power of the respective mean flow response in ideal hydrodynamics for LHC (grey) and RHIC (green) initial conditions. The bottom part of the plots shows the ratio of RHIC to LHC results at different opacities (different colors). The light grey band shows the error on the ratio of initial state ellipticity cumulants.  }
    \label{fig:cumulants_vs_centrality}
\end{figure*}

\subsection{Collective flow cumulants in OO collisions at RHIC and LHC}

Next, we discuss general properties of the statistics of final state elliptic flow results we found in our simulations, one of which will be important for the construction of new observables. Our results for the second and fourth order elliptic flow cumulants, as defined in Eqs. (\ref{eq:cumulant}) and (\ref{eq:epsp}), are presented in Fig.~\ref{fig:cumulants_vs_centrality}. In both cases we observe a smooth transition from the linear order flow response to the ideal hydrodynamic one as a function of shear viscosity.

The shape of the centrality dependence curves is a result of the competition of two effects. Less central events will have smaller opacity $\hat{\gamma}$, which decreases the flow response $\kappa$, but also larger initial state ellipticity $\epsilon_2$ which increases the final state elliptic flow $\varepsilon_p$. This results in a peak of the second order cumulant. The former of the two effects becomes stronger the larger the value of $\eta/s$, as the opacity dependence of $\kappa$ is steeper at smaller opacities, causing the position of the peak to shift as a function of $\eta/s$.

The fourth order cumulant becomes positive for less central events, indicating strong fluctuation contributions. The sign change of the fourth order cumulant happens almost consistently at the same centrality as determined by the geometry, while its magnitude varies strongly with the opacity. Flow responses at RHIC are slightly smaller than at LHC due to the smaller energy scale resulting in smaller opacities. When increasing the shear viscosity $\eta/s$, the ratio of RHIC to LHC results approaches unity, since the flow response curve $\kappa(\hat{\gamma})$ approaches a constant value in the large opacity limit. Comparing results for the second (fourth) order cumulant in the case of ideal hydrodynamics to cumulants of the initial state ellipticity $\epsilon_2$ multiplied by the respective squared (fourth power) mean flow response coefficient $\kappa_{\rm id}^2$ ($\kappa_{\rm id}^4$), we see remarkable agreement. We checked that this holds similarly also for the other simulation setups. In other words, it seems to be generally true that the following approximate equality holds:
\begin{align}
    \langle|\varepsilon_p|^n\rangle=\langle \kappa^n |\epsilon_2|^n\rangle \approx \langle{\kappa}\rangle^n \langle|\epsilon_2|^n\rangle\;.\label{eq:mean_kappa_dominance}
\end{align}

\begin{figure}
    \centering
    \includegraphics[width=.49\textwidth]{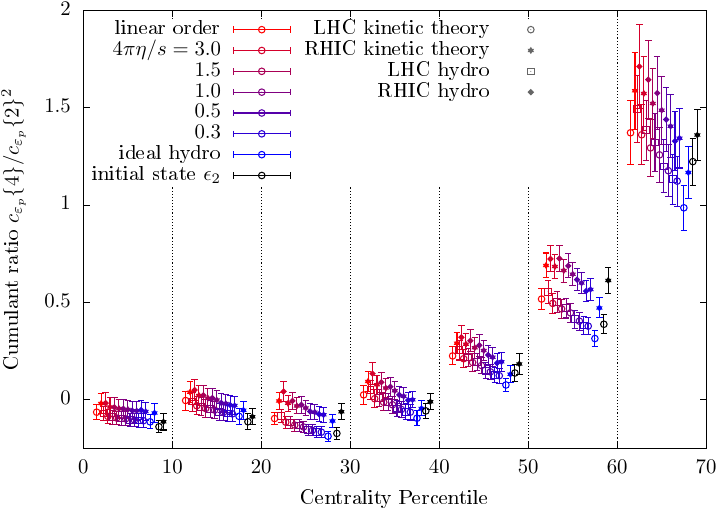}
    \caption{Ratio of fourth order cumulant to second order cumulant squared of the initial state ellipticity $\epsilon_2$ and of elliptic flow $\varepsilon_p$ at various values of the shear viscosity (different colors) as a function of centrality for simulations of OO collisions at LHC in kinetic theory (empty circles) and hydrodynamics (empty squares) as well as OO collisions at RHIC in kinetic theory (filled stars) and hydrodynamics (filled diamonds).}
    \label{fig:cumulant_ratio}
\end{figure}

\subsection{Constructing observables to disentangle response and geometry}\label{sec:deriving_observables}

 In the following, we discuss more formally the meaning of the finding in Eq.~\eqref{eq:mean_kappa_dominance} and then exploit it to construct our observables. Final state elliptic flow $\varepsilon_p$ can be interpreted as a dynamical response $\kappa$ to the initial state ellipticity $\epsilon_2$. A priori both the ellipticity and response will fluctuate event by event. We split the response into a mean and fluctuation:
\begin{align}
    \kappa&=\langle{\kappa}\rangle+\delta\kappa\;.
\end{align}
Accordingly, moments of the $\varepsilon_p$ distribution can be split into multiple contributions.
\begin{align}
    \langle\varepsilon_p^n\rangle=\sum_{k=0}^n \binom{n}{k}\langle{\kappa}\rangle^{n-k}\langle\delta\kappa^k\epsilon_2^n\rangle\label{eq:moment_splitting}
\end{align}
Depending on the ensemble that the above averages are taken over, the relative fluctuation of the response can be weak. As we have seen in the previous section, this is the case for the centrality classes of our simulation results. We expect that fluctuations in the response coefficient can be neglected whenever the fluctuation in opacity $\hat{\gamma}$ is weak. So the statement becomes more accurate for smaller centrality classes. If it is true, the single term in Eq.~\eqref{eq:moment_splitting} coming solely from the mean response to the initial state ellipticities dominates:
\begin{align}
    \langle\varepsilon_p^n\rangle \approx \langle{\kappa}\rangle^n\langle\epsilon_2^n\rangle\;.
    \label{eq:moment_approximation}
\end{align}
By definition, this approximation directly extends to the cumulants. This means that in certain ratios of the cumulants, the response can be eliminated~\cite{Bhalerao:2011yg} (to the extent that the approximation is valid):
\begin{align}
    c_{\varepsilon_p}\{2k\}/c_{\varepsilon_p}\{2\}^k\approx c_{\epsilon_2}\{2k\}/c_{\epsilon_2}\{2\}^k\;.\label{eq:cumulant_ratio}
\end{align}
This observation has already been made for $v_2\{6\}/v_2\{4\}$ and $\epsilon_2\{6\}/\epsilon_2\{4\}$~\cite{Giacalone:2016eyu}. In that case the authors mainly attribute the equivalence to the absence of a cubic response coefficient. For our purposes, we consider a possible such cubic coefficient to be a fluctuation of the response.

Now, to determine whether flow is generated due to a weakly or strongly interacting final state, we follow a similar idea to eliminate factors of the geometry by investigating the ratio between RHIC and LHC, i.e. between collision systems at two different energies. We exploit the fact that we can compare collision systems of the same or similar nuclei that should have very similar initial state geometry but slightly different opacity, which we will refer to as two ensembles "A" and "B". By taking the ratio of cumulants, the initial state geometry can then be approximately eliminated, 
\begin{align}
    c_{\varepsilon_p}\{2k\}|_A/c_{\varepsilon_p}\{2k\}|_B \approx \langle{\kappa}\rangle_A^{2k}/\langle{\kappa}\rangle_B^{2k}\;,
    \label{eq:cumulant_ratio_AB}
\end{align}
and the cumulant ratios yield ratios of powers of the response coefficients. However, we would like to be able to extract a value that can tell us on an absolute scale, how close a system is to hydrodynamic behaviour. Thus we propose to consider the logarithm of the ratio, which is the same as the difference in the logarithms. Dividing this by the difference in the logarithms of the mean opacity in the ensembles A and B, given that it is sufficiently small, we can approximate the logarithmic derivative of the logarithmic response:
\begin{align}
    \frac{\log(\langle{\kappa}\rangle_A/\langle{\kappa}\rangle_B)}{\log(\langle{\hat{\gamma}}\rangle_A/\langle{\hat{\gamma}}\rangle_B)}\approx \frac{\d \log(\kappa)}{\d \log(\hat{\gamma})}\left(\hat{\gamma}=\frac{\langle{\hat{\gamma}}\rangle_A+\langle{\hat{\gamma}}\rangle_B}{2}\right)\;.
\end{align}

If we want to extract this observable from experiment, we need to express $\langle{\hat{\gamma}}\rangle_A/\langle{\hat{\gamma}}\rangle_B$ in terms of measurable quantities. Since we consider ensembles with a similar geometry, the rms radii $R$ should also be similar. Furthermore, we can expect the specific shear viscosity to be comparable, if the systems are not too far apart in collisional energy. Thus, the ratio of opacities is given by the ratio of initial transverse energies, which we approximate by the ratio of final transverse energy per unit rapidity \footnote{We relax this approximation in App.~\ref{app:fwork_corr}.}
\begin{align}
    \frac{\langle{\hat{\gamma}}\rangle_A}{\langle{\hat{\gamma}}\rangle_B}&\approx\left(\left.\left\langle{\frac{\d E^{(0)}_\perp}{\d \eta}}\right\rangle_A\right/\left\langle{\frac{\d E^{(0)}_\perp}{\d \eta}}\right\rangle_B\right)^{1/4}\\
    &\approx\left(\left.\left\langle{\frac{\d E_\perp}{\d \eta}}\right\rangle_A\right/\left\langle{\frac{\d E_\perp}{\d \eta}}\right\rangle_B\right)^{1/4}\label{eq:Eratio_estimate}
\end{align}
Summarizing the discussion above, we can approximate the value of $\d \log\kappa / \d \log\hat{\gamma}$ in-between the two ensembles A and B by the following combination of experimental observables:
\begin{align}
      \W=\frac{2}{k} \cdot \frac{\log\left(c_{\varepsilon_p}\{2k\}|_A/c_{\varepsilon_p}\{2k\}|_B\right)}{\log\left(\left.\left\langle{\frac{\d E_\perp}{\d \eta}}\right\rangle_A\right/\left\langle{\frac{\d E_\perp}{\d \eta}}\right\rangle_B\right)}\approx\frac{\d \log\kappa}{\d \log \hat{\gamma}}\label{eq:W_def}
\end{align}
On the theory side, the function $\d \log\kappa / \d \log\hat{\gamma}$ computes the local power law exponent of the response curve $\kappa(\hat{\gamma})$, which, as we will explain now, contains important information on the degree of hydrodynamic behavior.
Namely, in accordance with our general discussion, for small opacities, the response is linear, while for large opacities, it saturates to a constant:
\begin{align}
    \kappa(\hat{\gamma}\ll 1)&=\kappa_{\rm LO}\hat{\gamma}\;,\\
    \kappa(\hat{\gamma}\gg 1)&=\kappa_{\rm id}\;.
\end{align}
Since the response curve $\kappa(\hat{\gamma})$ smoothly interpolates between these two limits, the quantity $\d \log\kappa / \d \log\hat{\gamma}$ monotonically transitions between the following limiting cases, informing our expectations for $\W$:
\begin{align}
    \W&= 0\; &&\mathrm{(Ideal\ Hydro)}\;,\\
\W&= 1\; &&\mathrm{(Noninteracting)}\;,
\end{align}
such that the quantity $\W$ provides a calibrated measure of the degree of hydrodynamic behavior.

We can further extract a theory calibration for $\W$ via $\d\log\kappa / \d\log\hat{\gamma}$ as computed from our fit of the flow response curve, which establishes a precise one-to-one correspondence between $\W$ and the mean opacity $(\langle\hat{\gamma}\rangle_A+\langle\hat{\gamma}\rangle_B)/2$ of the two systems. We note that for practical purposes, the inverse function computing $\hat{\gamma}$ from the observable $\W$ can be approximated quite well by the expression
\begin{align}
\hat{\gamma}\approx 2.5\frac{(1-\W)^{0.78}}{\W}
\end{align}
for values $0.05\lesssim \W \lesssim 0.9$.

\subsection{Validation and benchmark in event-by-event kinetic theory simulations}\label{sec:simulation_results}

We have made several approximations in the process of deriving the observable $\W$. How much measurements of this quantity can actually tell us about the considered collision systems will in practice depend on how accurately these approximate equalities hold. The approximations in Eqs.~\eqref{eq:cumulant_ratio_AB} and \eqref{eq:Eratio_estimate} require differences of key geometric properties to be negligible. In the following, we assess the interpretability of $\W$ using our simulation results. We also perform a first extraction of $\W$ from experimental data.

We first test the accuracy of the approximation via the dominance of the mean response [see Eq.~\eqref{eq:moment_approximation}] by verifying the cancellation of response coefficients in the ratios of different order cumulants as in Eq.~\eqref{eq:cumulant_ratio}. Fig.~\ref{fig:cumulant_ratio} shows the ratio of the fourth order cumulant to the second order cumulant squared $c\{4\}/c\{2\}^2$ for elliptic flow $\varepsilon_p$ and initial state ellipticity $\epsilon_2$ from all considered simulation setups (various values for $\eta / s$; various collision systems in hydro and kinetic theory). We observe that all results agree well with each other, confirming that the response cancels out. This means that these ratios carry direct information on the initial state geometry without sensitivity to the dynamical mechanism generating the flow response and can be used to validate or constrain initial state models or to investigate nuclear structure effects. Slight differences between RHIC and LHC can be observed in the cumulant ratio, indicating that the geometry is similar but not precisely identical. However, this observable characterizes a single collisional system and generally contains the same information in both cases separately.

\begin{figure}
    \centering
    \includegraphics[width=.49\textwidth]{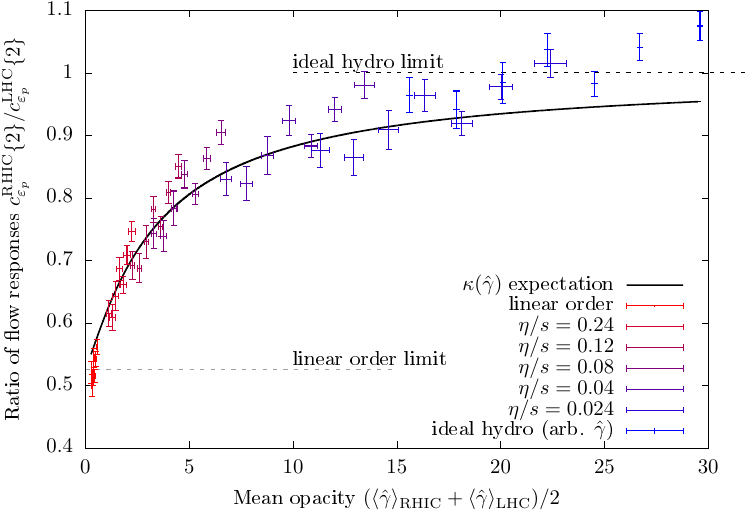}
    \caption{Ratio of cumulants $c_{\varepsilon_p}\{2\}$ between corresponding centrality classes of size 10\% of OO collisions at RHIC and LHC from our kinetic theory simulation results as a function of the mean opacity between both systems on a range in shear viscosity from the opacity-linearized case (red) to ideal hydrodynamics (blue), using various values of $\eta/s$ in-between (color gradient from red to blue), compared to the expectation from the $\kappa(\hat{\gamma})$-curve with the limits of linear order (light gray dashed) and ideal hydrodynamics (dark gray dashed). Error bars show the difference in opacity of the two systems and propagated error on the two cumulants.}
    \label{fig:kappa_ratio_vs_gamma}
\end{figure}

Next, we want to verify based on our simulation results that the observable in Eq.~\eqref{eq:W_def} can indeed serve as a measure of the interaction rate in the system. As a first step, we check whether the ratio of flow cumulants between two systems behaves like a power of the ratio of flow response coefficients in those two systems, i.e. if Eq.~\eqref{eq:cumulant_ratio_AB} holds. Figure~\ref{fig:kappa_ratio_vs_gamma} shows the ratio $c_{\varepsilon_p}^{\rm RHIC}\{2\}/c_{\varepsilon_p}^{\rm LHC}\{2\}$ for the centrality classes of our two OO systems as a function of the mean opacity of the two centrality classes. Simulation results at different shear viscosities are shown in different colors. This ratio is expected to behave like $\langle{\kappa}\rangle^2_{\rm RHIC}/\langle{\kappa}\rangle^2_{\rm LHC}$. At large opacities, the flow response saturates to the ideal hydrodynamic limit in both systems, so the ratio of response coefficients should go to 1 as indicated by the upper gray dashed line. At small opacities, flow responses should be linear in the opacity, such that $\langle{\kappa}\rangle^2_{\rm RHIC}/\langle{\kappa}\rangle^2_{\rm LHC}=\langle{\hat{\gamma}}\rangle^2_{\rm RHIC}/\langle{\hat{\gamma}}\rangle^2_{\rm LHC}$ as indicated by the lower gray dashed line. The black line shows as a theory expectation the squared ratio of the flow response coefficients in the two systems as estimated by the flow response curve, i.e. $\kappa^2(\langle\hat{\gamma}\rangle_{\rm RHIC}) / \kappa^2(\langle\hat{\gamma}\rangle_{\rm LHC})$.

Indeed, our linear order and ideal hydrodynamic results are in line with these expectations. Furthermore, results from finite $\eta/s$ smoothly interpolate between these two limits, following the theory expectation curve, shown in black. This means that the ratio of flow cumulants from two different systems may indeed serve as a proxy for the ratio of flow responses in these systems. There is some fluctuation in between centrality classes. Some of it is due to the fact that the ratio of mean opacities varies slightly for different centrality classes. But even in the ideal hydrodynamic case the response coefficients are not exactly the same between centrality classes. Still, within error bars, the results can be said to follow the curve corresponding to the theoretical expectation.

\begin{figure*}
    \centering
    \includegraphics[width=.49\textwidth]{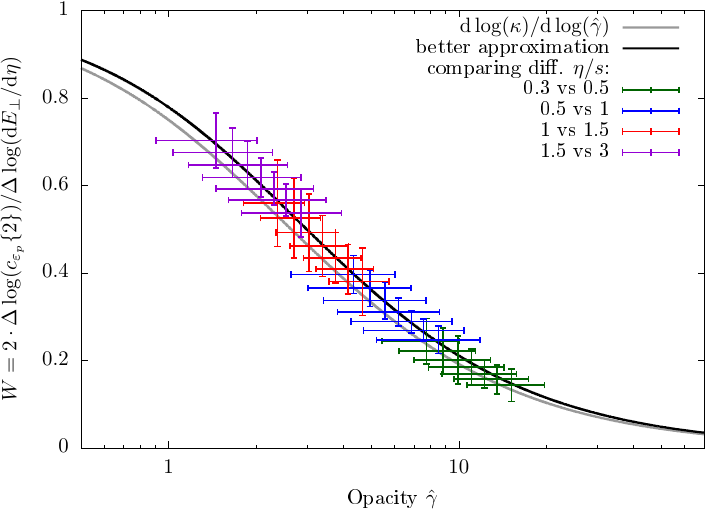}
    \includegraphics[width=.49\textwidth]{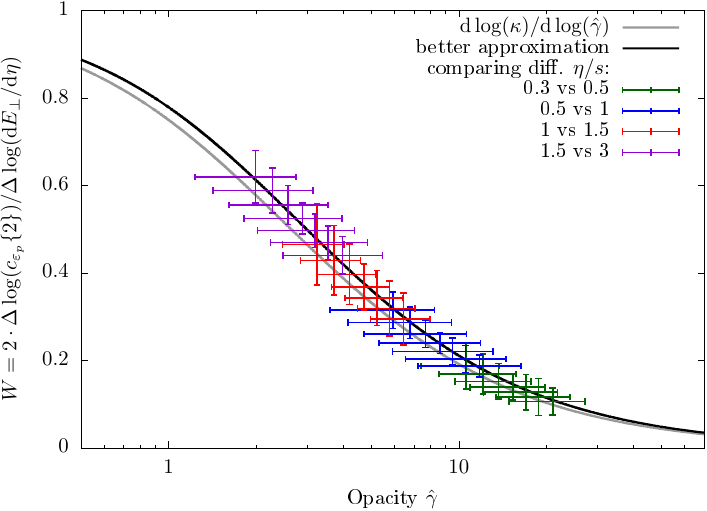}
    \includegraphics[width=.49\textwidth]{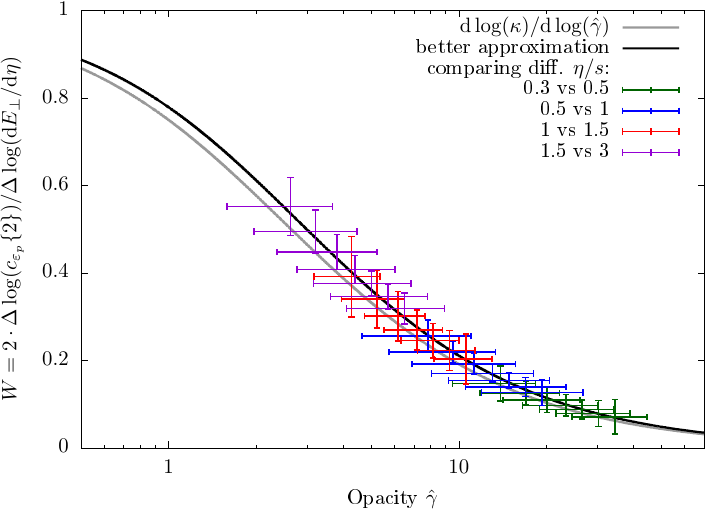}
    \includegraphics[width=.49\textwidth]{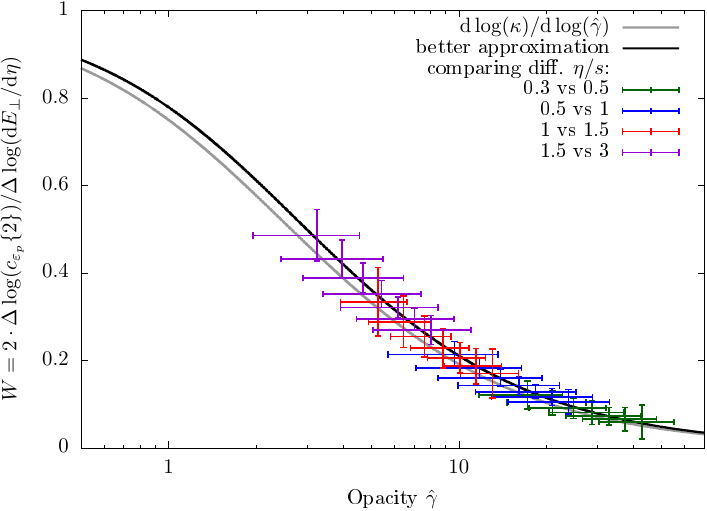}
    \caption{
    \label{fig:log_flow_curve_etas}    
    Kinetic theory simulation results from centrality classes of size 10\% up to the 60-70\% class of OO collisions at RHIC (top left) and LHC (top right), as well as AuAu collisions (bottom left) and PbPb collisions (bottom right) for $\W$ computed from pairs of ensembles with exactly the same geometry but differing shear viscosities $\eta/s=0.024$ and $0.04$ (green), $\eta/s=0.04$ and $0.08$ (blue), $\eta/s=0.08$ and $0.12$ (red), as well as $\eta/s=0.12$ and $0.24$ (purple), where the transverse energy was scaled by $(\eta/s)^{-4}$ in accordance with the definition of the opacity~\eqref{eq:ghat}, compared to the theory expectation $\d\log\kappa / \d\log\hat{\gamma}$ (grey) and the improved calibration curve (black). Error bars show the difference in opacity between the two systems and the propagated error of the cumulants and transverse energies.}
\end{figure*}

We move on to studying directly the behavior of the $\W$ observable. As an idealized test of its capability to quantify the interaction rate in a system, we compute it from two simulation results with exactly the same initial state ensemble but at different shear viscosity. The construction of $\W$ assumes that the shear viscosity in the two systems is almost the same, such that the ratio in opacities is given by the ratio of transverse energies. Thus, for this test, in order to use the same definition of $\W$ but correctly describe the opacity dependence, we multiply the values of the transverse energy in each case by $(\eta/s)^{-4}$ [c.f. Eq.~\eqref{eq:ghat}]. Fig.~\ref{fig:log_flow_curve_etas} shows results for $\W$ obtained for OO collisions at RHIC in the upper left panel and at LHC in the upper right panel, as well as AuAu collisions in the lower left panel and PbPb collisions in the lower right panel.  $\W$ was computed for centrality classes of size 10\% up to the 60-70\% class from simulations with different pairs of $\eta/s$ values plotted in different colors. They are plotted as a function of the mean opacity between the two simulation setups, where the error is given by the difference in opacity. These results for $\W$ are compared to the theory calibration curve $\d\log(\kappa) / \d\log(\hat{\gamma})$ in gray, as well as an improved version of this curve in black, which also takes into account the difference between the initial state and final state ratio of transverse energies (see App.~\ref{app:fwork_corr}).

Focusing now on our extracted values of $\W$, they do very closely follow the calibration curves for all collision systems and all pairs of shear viscosity values. Different collision systems and shear viscosities have a different range in opacity, such that we effectively scanned the opacity range $\hat{\gamma}\sim 1.5-40$. Thus we can conclude that $\W$ does behave as expected. The agreement is slightly better with the improved calibration curve; however, given the error of our $\W$ results, both theory curves are compatible.

\begin{figure}
    \centering
    \includegraphics[width=.49\textwidth]{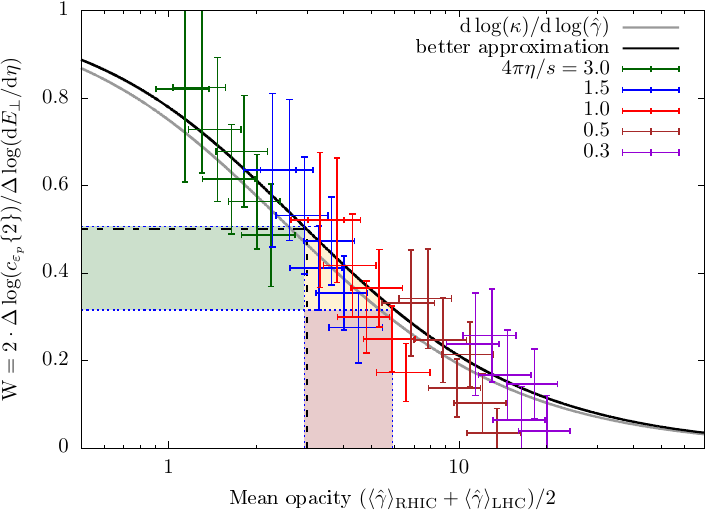}
    \caption{$\W$ observable obtained for the centrality classes of our simulations of OO collisions at LHC and RHIC energies (crosses) at different values of the shear viscosity (different colors). The results agree well with the calibration curve from the theory side, where the grey line represents the approximation in Eq. (\ref{eq:W_def}) and the black line shows the improved result from Eq. (\ref{eq:calibration_improved}). Dashed lines and colored areas demonstrate how measured ranges in $\W$ can be turned into ranges for the predicted mean opacity. }
    \label{fig:log_flow_curve}
\end{figure}

Due to the $\sqrt{s}$-dependence of the nucleon-nucleon scattering cross-section, in reality, the collision geometry between the two ensembles that enter the calculation of $\W$ will not be exactly the same, even if we compare collisions of the same nuclei at slightly different energies $\sqrt{s_{NN}}$ and it should thus be tested how much this affects the observable.
To this end, we also computed $\W$ by combining results from the two OO collision systems, as well as from PbPb and AuAu. Fig.~\ref{fig:log_flow_curve} compares these values for the two OO systems to the extracted calibration curve of ${\d\log(\kappa)}/{\d\log(\hat{\gamma})}$ and the improved approximation of $\W$ (see App.~\ref{app:fwork_corr}). Within the level of accuracy that we were able to determine it, the results closely follow the curve. The error on $\W$ would be decreased simply by increasing the sample size. The error on $\hat{\gamma}$ is given by the difference in opacity between LHC and RHIC. To some degree,  considering collision systems that are closer in opacity would improve the accuracy; however, there is a limit to it. The approximations that we made in order to establish the correspondence of $\W$ and ${\d\log \kappa}/{\d\log \hat{\gamma}}$ in Eq. (\ref{eq:W_def}) are valid as long as the effect of event-by-event fluctuations of the flow response is negligible compared to the difference in the average response between the two systems.

\begin{figure}
    \centering
    \includegraphics[width=.49\textwidth]{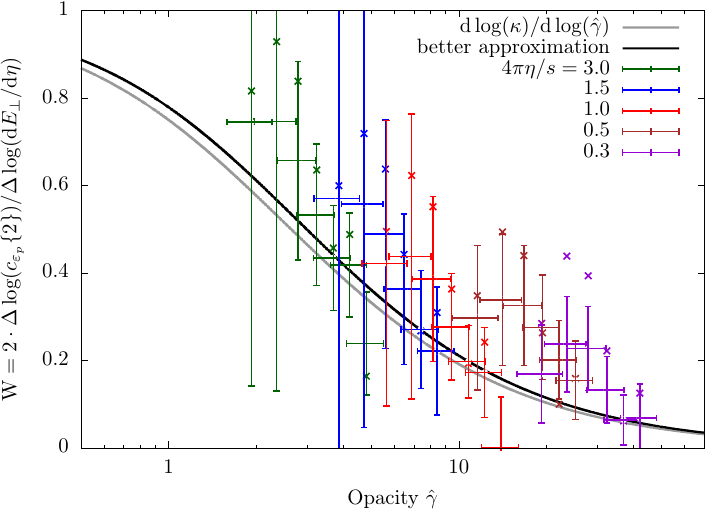}
    \caption{$\W$ computed from kinetic theory simulation results of centrality classes of size 10\% up to the 60-70\% class for the PbPb and AuAu collision systems at different shear viscosities $\eta/s$ in different colors, comparing results by definition~\eqref{eq:W_def} of $\W$ (crosses) and the modification~\eqref{eq:W_mod} by rescaling by initial ellipticity cumulants (pluses). Error bars show the difference in opacity between the two systems and the propagated error of the cumulants and transverse energies for the modified quantity.
    The results are compared to  ${\d\log(\kappa)} / {\d\log(\hat{\gamma})}$ (gray) and the improved theory curve (black).
    }
    \label{fig:log_flow_curve_RHICvLHC}
\end{figure}

\begin{figure}
    \centering
    \includegraphics[width=.49\textwidth]{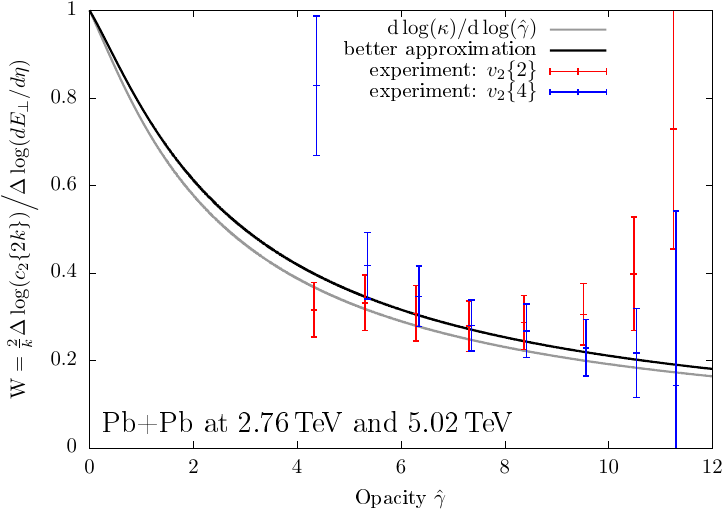}
    \caption{$\W$ computed from experimental data for $v_2\{2\}$ (red) and $v_2\{4\}$~\cite{ALICE:2018rtz} (blue) as well as transverse energy~\cite{ALICE:2022imr} for centrality classes for centrality classes 0-5\%, 5-10\%, 10-20\%, 20-30\%, 30-40\%, 40-50\%, 50-60\% and 60-70\% in PbPb collisions at $2.76$ and $5.02\,$TeV as a function of the \trento ~estimate for the mean opacity between the two systems, with $\eta/s=0.12$ taken as the value at which we reproduce experimental data for transverse energies. Error bars show the propagated error of the experimental data. 
    The results are compared to  ${\d\log(\kappa)} / {\d\log(\hat{\gamma})}$ (gray) and the improved theory curve (black).
    }
    \label{fig:log_flow_curve_LHC_exp}
\end{figure}

Having established Eq. (\ref{eq:W_def}), the calibration curve can be used to translate measurements of $\W$ into mean opacity values of the considered centrality class in LHC and RHIC collisions. The blue dotted lines and colored area in Fig.~\ref{fig:log_flow_curve} show how the extracted $\W$ range for the 30-40\% centrality class at $4\pi\eta/s=1.5$ turns into a range for this mean opacity. The black dashed line shows the limit of applicability of hydrodynamics at $\hat{\gamma}=3$ as established in our previous work~\cite{Ambrus:2022koq,Ambrus:2022qya}, which corresponds to $\W=0.5$. Some other values are given in Table~\ref{tab:W_values}, while the extracted estimates for the mean opacities that we obtained from our results for $\W$ are presented in Table~\ref{tab:ghat_determination}. The form of the calibration curve makes it harder to find accurate values for smaller opacities. In order to achieve a given relative uncertainty in $\hat{\gamma}$, roughly the same amount of relative uncertainty in $\W$ is required at large opacities $\hat{\gamma}\gtrsim 10$, whereas at small opacities $\hat{\gamma}\sim 1$ a relative uncertainty which is five times smaller is necessary.

\begin{table}[]
    \centering
    \begin{tabular}{c|ccccccc}
        $\hat{\gamma}$ & 0.5 & 1.0 & 3.0 & 4.0 & 6.0 & 9.0 & 12.0 \\
        \hline
       $\W$ & 0.89 & 0.78 & 0.50 & 0.42 & 0.32 & 0.23 & 0.18
    \end{tabular}
    \caption{Example values of the calibration curve linking opacity $\hat{\gamma}$ and the observable $\W$, computed according to Eq. (\ref{eq:calibration_improved}).}
    \label{tab:W_values}
\end{table}

\begin{table}[]
    \centering
    \begin{tabular}{c|ccccc}
    Centrality& \multicolumn{2}{c}{actual $\hat{\gamma}$}&\multicolumn{3}{c}{mean $\hat{\gamma}$}\\
    class & RHIC & LHC & estimates\\
    \hline\hline
         0-10\%&3.75(19)&5.20(25)& $7.22^{+3.11}_{-1.91}$\\
        10-20\%&3.35(13)&4.66(18)& $5.13^{+1.74}_{-1.19}$\\
        20-30\%&3.04(12)&4.21(16)& $3.30^{+1.06}_{-0.78}$\\
        30-40\%&2.74(12)&3.80(16)& $4.13^{+1.34}_{-0.95}$\\
        40-50\%&2.46(11)&3.39(16)& $2.69^{+1.18}_{-0.83}$\\
        50-60\%&2.18(11)&2.99(15)& $1.84^{+0.99}_{-0.71}$\\
        60-70\%&1.91(10)&2.61(14)& $1.84^{+1.14}_{-0.79}$
    \end{tabular}
    \caption{Comparison of mean opacity values computed from initial profiles of several centrality classes of OO collisions at RHIC and LHC with the estimates obtained by plugging results for the observable $\W$ from simulations at a shear viscosity of $4\pi\eta/s=1.5$ into the calibration curve given in Eq.~\eqref{eq:calibration_improved}.}
    \label{tab:ghat_determination}
\end{table}

We also check whether $\W$ can still be accurately extracted by combining results from colliding nuclei that are similar in structure, but not identical. In Fig.~\ref{fig:log_flow_curve_RHICvLHC}, the presented results for $\W$ were computed by combining PbPb and AuAu results at the same shear viscosity and of the same centrality class, for centrality classes of size 10\% up to the 60-70\% class. Different colors show simulation results at different values of the shear viscosity. The crosses show results from unadulterated data of both systems. They significantly overshoot the calibration curves. It seems that when comparing PbPb and AuAu, the slight differences in geometry are still too large for an accurate extraction of $\W$. 
Note that the geometries corresponding to the RHIC and LHC energies for the OO system were very similar, as shown by the eccentricity cumulants $c_{\epsilon_2}\{2\}$ shown as gray and green bands in the left panel of Fig.~\ref{fig:cumulant_ratio}.
However, this is not the case for the PbPb (LHC) and AuAu (RHIC) systems. There, the corresponding cumulants $c_{\epsilon_2}\{2\}$ differ by approximately $10\%$, as can be seen in Figs.~6 and 7 in the companion paper~\cite{Ambrus:2024hks}.
Thus, we also computed a version of $\W$ where the ratio of initial ellipticity cumulants between the two systems is taken into account, i.e. we compute it as
\begin{align}
      \W_{\rm mod}&=2 \cdot \frac{\log\left(\frac{c_{\varepsilon_p}\{2\}|_{\rm RHIC}}{c_{\varepsilon_p}\{2\}|_{\rm LHC}}\frac{c_{\epsilon_2}\{2\}|_{\rm LHC}}{ c_{\epsilon_2}\{2\}|_{\rm RHIC}} \right)}{\log\left(\left.\left\langle{\frac{\d E_\perp}{\d \eta}}\right\rangle_{\rm RHIC}\right/\left\langle{\frac{\d E_\perp}{\d \eta}}\right\rangle_{\rm LHC}\right)}\label{eq:W_mod}\;.
\end{align}
With this modification, the values for $\W_{\rm mod}$ are now consistent with the calibration curves. However, this quantity now has significantly larger errors. Of course, modifying $\W$ in this way defeats its purpose, because it brings back the necessity of modeling the initial state geometry if one wants to extract it from experimental data. The lessons one might take from this are twofold. If one wants a theory independent determination of $\W$, one should compare experimental data from collisions of the same nuclei, where geometrical differences in the initial state should be minimal.
This was indeed the case in our OO setup, where we considered just the variation of the nucleon-nucleon cross-section with collision energy. Note, however, that taking into account other factors, such as overall nucleon size or nuclear density profiles, could lead to more significant geometry discrepancies (see, e.g., the IP-Glasma analyses of Refs.~\cite{Singh:2023rkg,Mantysaari:2024qmt}).

We also point out that relative errors can become large when considering error propagation through a logarithm. This means that achieving a reasonable size of the error on $\W$ requires significantly more precise measurements of the quantities that go into its determination the closer the $\sqrt{s_{NN}}$ of the two considered systems, since the logarithm of a ratio close to 1 will be close to 0, making relative errors large.

\subsection{Comparison to experimental data in PbPb collisions}
Even though the primary purpose of the observable $\W$ is to quantify the degree of hydrodynamic behavior in small systems, it is tempting to exploit existing LHC measurements of PbPb collisions at  $2.76\;$TeV and $5.02\;$TeV. We present this analysis in Fig.~\ref{fig:log_flow_curve_LHC_exp}, which shows an extraction of $\W$ from experimental data of PbPb collisions at $2.76\;$TeV and $5.02\;$TeV~\cite{ALICE:2018rtz,ALICE:2022imr} compared to the calibration curves. The corresponding opacity values were computed from our \trento~ensembles with a shear viscosity value of $\eta/s=0.12$, which yields simulation results for transverse energies that are compatible with experiment. We computed two versions of $\W$: one from $v_2\{2\}$ data plotted in red and the other one from $v_2\{4\}$ data plotted in blue, each for centrality classes 0-5\%, 5-10\%, 10-20\%, 20-30\%, 30-40\%, 40-50\%, 50-60\% and 60-70\%. Since the experiments only provide results for particle number weighted flow harmonics, we applied a scheme devised in~\cite{Kurkela:2019kip} to convert them into energy weighted flow harmonics, where  particle number weighted results for $v_2$ were multiplied by $1.33$ in the $2.76\;$TeV case and by $1.34$ in the $5.02\;$TeV case. Clearly, in the ratio between the two systems, this is not a large difference;  however for the result for $\W$, i.e. after taking the logarithm of this ratio, the change due to this rescaling was noticeable, and it is needless to say that a genuine experimental extraction of energy weighted flow harmonics would be preferable. When comparing our extraction of $\W$ to the calibration curve, we find that all rescaled results except the most central one for $v_2\{2\}$ and the most peripheral one for $v_2\{4\}$ are compatible with the calibration curves. It might seem that the centrality dependence is off in the $v_2\{2\}$ case. This might be due to unwanted nonflow contributions, but the size of the error bars also allows to attribute it to statistics. In the $v_2\{4\}$ case, the results clearly show the expected centrality dependence. This first test shows that $\W$ can also in experimental practice provide a meaningful quantification of the interaction rate in a given system.

We want to make one further comment regarding the experimental extraction. Since we only found compatible data directly for $v_2\{2k\}$, we used this instead of the cumulants. The $v_2\{4\}$-based result has particularly large error bars in the most central and most peripheral result, because from the experimental side no data were given in cases where the flow cumulant had the wrong sign for $v_2\{4\}$ to be extracted. For our scheme, when plugging in the cumulant directly, the sign does not matter as it should be dictated by the sign of the corresponding geometry cumulant and thus be the same in both cases and cancel in the ratio.

\section{Hydrodynamization observable for a nonconformal system}\label{sec:non-conformal}

Clearly, one of the shortcomings of validating the $\W$ observable in our model description is that it is conformal. Even though the motivation for $\W$ is rather general and expected to hold away from conformality, nonconformal effects may change the quantitative phenomenology. We tried to probe at least some of the effects that deviation from a perfectly conformal system can have and we will discuss the results in this section.

We chose the most straightforward way of moving the description away from conformality closer towards real QCD that was available to us, which was to simply run hydrodynamic simulations in the code vHLLE~\cite{Karpenko:2013wva} with both a conformal equation of state and the QCD equation of state~\cite{Steinheimer:2010ib} that comes with it. We apply a centrality class dependent global scaling factor to the initial profile in order to make the final state results for transverse energy fit with those obtained from kinetic theory at $\eta/s=0.12$, which is in good agreement with measurements in PbPb collisions. We refer to our companion paper~\cite{Ambrus:2024hks} for more details on the implementation, and now move on to discuss the results and compare to conformal hydrodynamics and kinetic theory.

\begin{figure}
    \centering
    \includegraphics[width=0.99\linewidth]{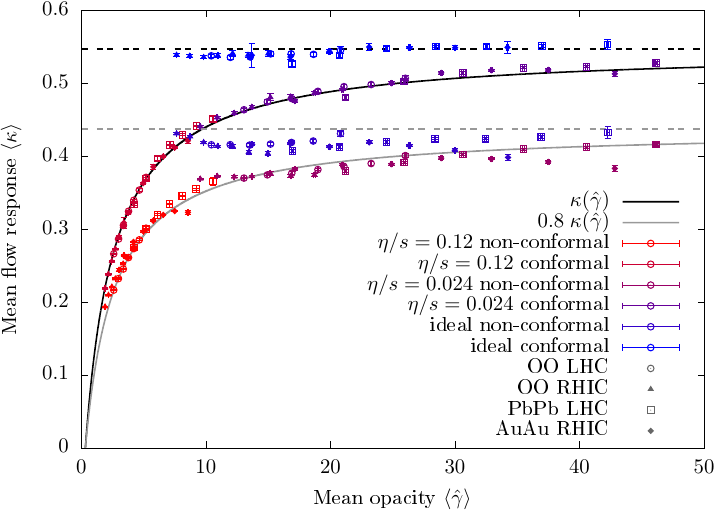}
    \caption{Mean values of the response coefficient $\kappa=\varepsilon_p/\epsilon_2$ as a function of the mean opacity 
    (this is chosen arbitrarily for ideal hydrodynamics)
    for centrality classes of size 10\% from conformal kinetic theory and nonconformal hydrodynamic (different colors) simulations of OO collisions at LHC (empty circles) and RHIC (filled triangles) as well as PbPb collisions (empty squares) and AuAu collisions (filled diamonds) at different values of the shear viscosity (different colors) compared to the conformal kinetic theory flow response curve $\kappa(\hat{\gamma}$ (black solid) and its ideal hydrodynamic limit (black dashed) as well as a rescaling of the same curve by a factor of $0.8$ (grey solid and dashed).}
    \label{fig:kappa_fit_switch}
\end{figure}

We first assess how the nonconformal results compare on the level of the flow response $\kappa(\hat{\gamma})$ in Figure~\ref{fig:kappa_fit_switch} , where we compare nonconformal flow response results to conformal ones. Different symbols indicate different collision systems and different dynamical models are shown in different colors. The black line shows the flow response curve $\kappa(\hat{\gamma})$ as obtained from the conformal hydrodynamic simulation data and the black dashed line shows the ideal hydrodynamic limit. Grey solid and dashed lines show the conformal response curve scaled by a constant factor of $0.8$, which approximately describes the nonconformal response. 

We note that, if the scaled $\kappa(\hat{\gamma})$-curve would perfectly describe also the nonconformal data, then the calibration curve for $\W$ would work just the same in this case, as it is insensitive to a constant prefactor. However, it seems that already the ideal hydrodynamic results fall slightly below the constant scaling expectation. When including viscous effects by considering a finite $\eta/s$, we find that the LHC data are actually described quite well by the rescaled conformal response curve. However, at RHIC energies the nonconformal equation of state gives rise to an additional centrality dependence, such that RHIC results for the response coefficient fall below the scaled conformal curve for central classes and above the curve for peripheral classes. Since opacity scales are different for different systems, this behavior cannot be captured by a single opacity dependent curve. While, on a global scale, the deviation from the scaled curve does not seem to be that strong, the $\W$ observable is sensitive to differences in flow responses, so these local deviations play a bigger role in its extraction. However, there are ways to alleviate the problem, as will be discussed in the following.

\begin{figure*}
    \centering
    \includegraphics[width=.49\linewidth]{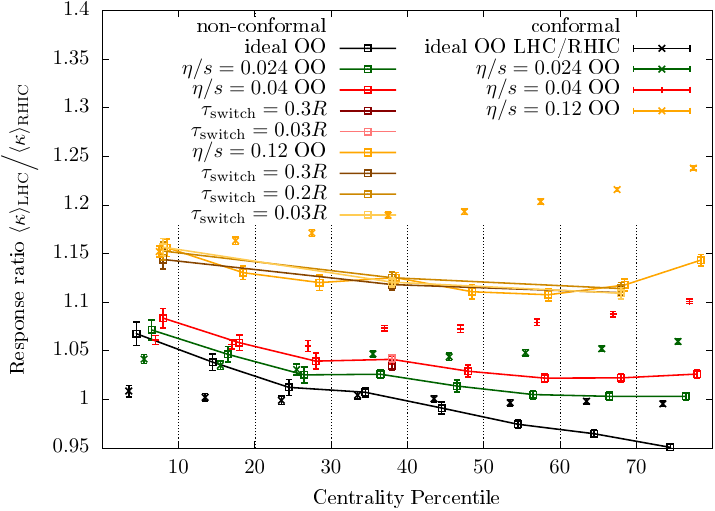}
    \includegraphics[width=.49\linewidth]{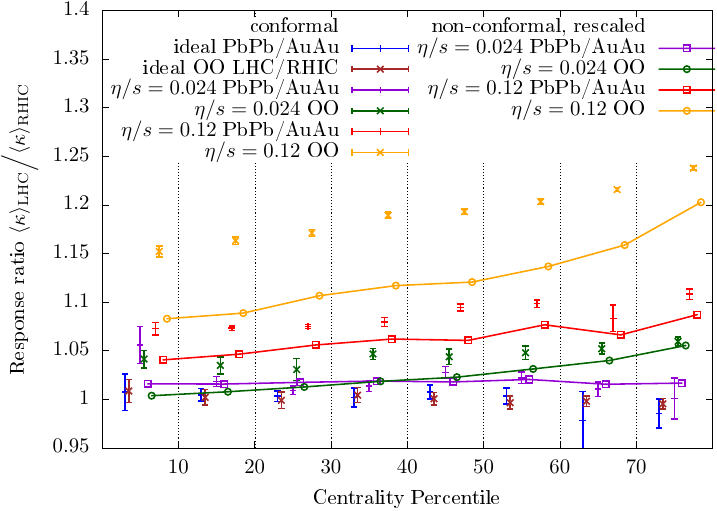}
    \caption{Ratios of mean flow response coefficients $\kappa=\varepsilon_p/\epsilon_2$ between RHIC and LHC collision systems as a function of centrality classes for conformal (crosses) and nonconformal simulation results (empty squares with lines). Different colors indicate different setups, where we varied the collision system (two OO systems or PbPb vs AuAu), the shear viscosity $\eta/s$ or the time that the eos is switched. In the right plot nonconformal results at finite $\eta/s$ were scaled by the ratio of ideal hydrodynamic results at RHIC and LHC as a function of centrality. Error bars show the propagated errors of the cumulants.}
    \label{fig:kappa_ratio_switch}
\end{figure*}

Whether the deviating behavior affects the interpretability of $\W$ can be checked most straightforwardly by comparing the ratio of flow response coefficients between RHIC and LHC in the conformal and nonconformal cases. Fig.~\ref{fig:kappa_ratio_switch} shows results for the ratio of mean values of $\varepsilon_p/\epsilon_2$ between RHIC and LHC in corresponding centrality classes of size 10\% up to the 70-80\% class. Results for different collision systems and/or values of the shear viscosity are plotted in different colors. Conformal results are plotted as crosses or pluses with error bars, while nonconformal results are plotted as open symbols. The left plot shows unadultered results and we also indicate extra points for the nonconformal case of $\eta/s=0.04$ and $\eta/s=0.12$, where we varied the time at which the equation of state is switched during the evolution.

Clearly, the conformal results indicate the behavior that is expected and necessary for the conceptualization of $\W$. Namely, in the ideal case, the ratio shows no centrality dependence because the flow response has no opacity dependence, whereas with decreasing overall opacity, the slope of the ratio increases. This is due to the fact that the differences in the ratio of LHC to RHIC opacities cause a larger difference in the flow response the smaller the mean opacities in the respective centrality class. The left plot shows that the feature of an increasing slope is still observed in the nonconformal case. However, due to the additional centrality dependence of nonconformal effects, already at the level of ideal hydrodynamics one observes a decrease of the ratio as a function of centrality. While at LHC the nonconformal flow results are compatible with a constant scaling of the conformal results, at RHIC they undershoot the scaled conformal results in central classes and overshoot them in peripheral classes, causing the observed behavior of the ratio.

Since the behavior in ideal hydrodynamics provides the baseline for assessing the magnitude of dissipative effects, the offset baseline of the ratio could be cured by focusing only on mid-central classes. Alternatively, one can try to cancel it out by rescaling nonconformal results at finite $\eta/s$ by the ratio of conformal to nonconformal results in the ideal case for the respective centrality class, as depicted in the right panel of Fig.~\ref{fig:kappa_ratio_switch}. Indeed now all nonconformal curves also have positive slopes as a function of centrality. While for smaller values of $\eta/s$ the nonconformal results are now in agreement with the conformal ones, the agreement deteriorates at larger $\eta/s$, indicating that for large opacities the  calibration curve for $\W$ needs to be readjusted to account for nonconformal effects.

Since it is theoretically difficult to capture the effects of a nonconformal equation of state in the nonequilibrium regime, one may be worried about a loss of theoretical control. By
going back to the left plot, the results from varying the switching time serve as an estimate of the level of theoretical control  that can still be achieved when switching the equation of state on the fly from a conformal preequilibrium evolution to a QCD equation of state in hydrodynamics. As it turns out, on the level of the ratio, the effect of varying the switching time is quite small. However, as shown in Appendix C of our companion paper~\cite{Ambrus:2024hks}, the time evolution in a sample simulation shows that this might be deceiving, as flow results in both LHC and RHIC results deviate quite a bit when varying the switching time, but in the same direction.

Despite these ambiguities related to nonconformal effects, we will now demonstrate that $\W$ can still provide meaningful information on the degree of hydrodynamization. In order to do so, we extracted $\W$ from the nonconformal simulation data and compare this to our theoretical expectations. To reflect the theory uncertainty in how to treat the response, we computed multiple Padé fits to the nonconformal flow results. Technically, in hydrodynamics, expectations on the flow response curve differ from kinetic theory in 
an unphysical way,
as discussed in App.~\ref{app:hydro_curves}. Thus, we choose two different ansätze for the fits to nonconformal data. The first is of the same form as the one we used in Sec.~\ref{sec:validaiton} for our kinetic theory results. For the other fitting function, we chose the following form.

\begin{align*}
    \kappa_{\rm switch}(\hat{\gamma})=\frac{\kappa_{0,\rm switch}+\kappa_{\rm LO,switch}\hat{\gamma}}{1+\frac{\kappa_{\rm LO,switch}}{\kappa_{\rm id,switch}}\hat{\gamma}}
\end{align*}

The ideal coefficient was obtained from the ideal hydrodynamic results as $\kappa_{\rm id,switch}=0.419$, while the other coefficients were extracted by the fitting procedure. We obtained three sets of parameters for each Padé curve, where we fitted them to only the LHC results, only the RHIC results or to all results together. This allows to reflect the theoretical uncertainty in the setup as a band of different conceivable theory calibration curves.

\begin{figure}
    \centering
    \includegraphics[width=0.99\linewidth]{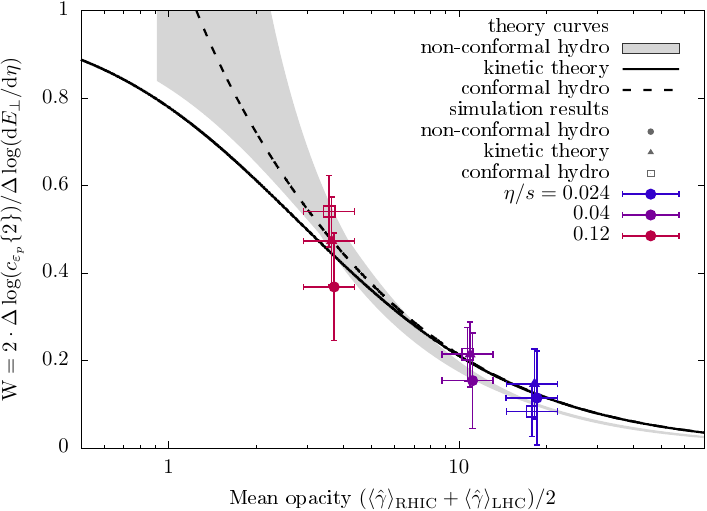}
    \caption{Hydrodynamization observable $\W$ computed from simulation data of the 20-30\% centrality class of OO collisions at RHIC and LHC as a function of the mean opacity between the two systems from kinetic theory (filled triangles) as well as conformal (empty squares) and nonconformal hydrodynamics (filled circles) for shear viscosities $\eta/s=0.024$ (blue), $0.04$ (purple) and $0.12$ (red), compared to the kinetic theory (black solid) and conformal hydrodynamic (black dashed) calibration curves as well as a range of possible calibration curves for nonconformal hydrodynamics (gray band). Error bars show the propagated error of cumulants and transverse energies. }
    \label{fig:kappa_W_switch}
\end{figure}

Figure~\ref{fig:kappa_W_switch} now compares results for $\W$ from our conformal and nonconformal simulations from the 20-30\% centrality classes of OO collisions at RHIC and LHC. The black solid line is our calibration curve from section~\ref{sec:simulation_results}, while the black dashed line shows an analogous curve for the case of our conformal hydrodynamic results (c.f. App.~\ref{app:hydro_curves}). The grey band shows the spread of the different calibration curves obtained from nonconformal results.
We find that the nonconformal data points for the $20-30\%$ centrality class indeed show the expected trend as a function of opacity, noting however that this would not have been the case if we had plotted also results from other centrality classes without scaling with the ideal response ratio, as was discussed earlier in this section. By comparing the individual data points one to one, we see that the nonconformal results for $\W$ have slightly lower values than conformal ones, which is also reflected in the behavior of the calibration curves. This means that it should still be possible to infer the degree of hydrodynamic behavior in the nonconformal system by comparing to an appropriate theory expectation.

In conclusion, our first simple test of nonconformal effects on $\W$ indicates that it should still be a feasible measure of hydrodynamization. In order to make reliable statements also at RHIC energies about whether a system close to $\W\sim 0.5$ should be considered as behaving hydrodynamically, further studies will have to be done to extract a more sound theory calibration curve. In any case, restricting the extraction to only one mid-central class, we found that $\W$ will give the correct ordering of different systems in terms of their interaction rate. Also, at sufficiently high collision energies $\sqrt{s_{NN}}$, e.g. at LHC, our results indicate that the nonconformal flow response can be assumed to be a constant rescaling of the conformal result, in which case $\W$ remains the same and the calibration curve obtained in our model could be consulted. We concede, however, that this first test of course does not rule out that nonconformal effects from other sources than the equation of state, which we did not consider here, might also affect the behavior of $\W$.

\section{Conclusion}\label{sec:conclusion}
We analyzed the behavior of energy weighted flow cumulants in OO collisions and found that for cumulants of the usual centrality classes, fluctuations of flow response coefficients can be neglected. Based on this approximation, we have constructed observables that disentangle the effects of the initial state geometry and the dynamical response mechanism on final state anisotropic flow. Ratios of different cumulants from the same collision system are essentially only sensitive to fluctuations of the geometry and can thus be used to probe e.g. nuclear structure effects in light ions~\cite{Zhang:2024vkh,Giacalone:2024luz,Giacalone:2024ixe}. Similarly, ratios of the same cumulant for collision systems of the same nuclei at different energies reflect the ratio of mean response coefficients between the two systems. Building on this observation, we exploited the fact that there is a universal flow response curve $\kappa(\hat{\gamma})$ in order to construct an experimentally accessible observable 
$$
 \W=\frac{2}{k} \cdot \frac{\log\left(c_{\varepsilon_p}\{2k\}|_A/c_{\varepsilon_p}\{2k\}|_B\right)}{\log\left(\left.\left\langle{\frac{\d E_\perp}{\d \eta}}\right\rangle_A\right/\left\langle{\frac{\d E_\perp}{\d \eta}}\right\rangle_B\right)}
$$
that is sensitive to the interaction rate in the system on an absolute scale. Since the initial state eccentricity is almost identical in the two systems, the ratios inside the logarithms can quantify two effects. The numerator measures the change in flow response, while information on the ratio of opacities between the two systems can be accessed via the ratio of final state energies in the denominator. Both ratios can be turned into differences via the logarithm function, such that the ratio of the two logarithms is then an approximation to the logarithmic derivative of the collective flow response, ${\d\log(\kappa)}/{\d\log(\hat{\gamma})}$, and measures the local power law of the flow response curve.

We argued on general grounds that this function has limiting values of 1 in the dilute limit and 0 in the ideal hydrodynamic limit. In between, it provides a smooth monotonic interpolation of these cases, meaning that it establishes a one-to-one correspondence between $\W$ and the average interaction rate of the two systems. Thus, $\W$ can be used as a measure of the degree of hydrodynamization. Its precise interpretation can be informed by theory calibration curves. Our own findings indicate that a system can be considered hydrodynamic if $\W\lesssim 0.5$.

We validated in simulations based on conformal RTA kinetic theory that if computed from cumulants and transverse energies, $\W$ indeed follows the expected behavior of the theory calibration curve. When computing it for centrality classes of two systems with exactly the same initial state geometry but different opacity scale, we found an accurate verification of this expectation. In the realistic case of comparing two systems with slightly different geometry, we found that for OO collisions at RHIC and LHC, $\W$ still provides accurate information on the mean interaction rate. However, when computed from AuAu and PbPb simulations at RHIC and LHC, the small differences in the initial state geometry already spoil the interpretability of the $\W$ observable. This can be alleviated by modifying $\W$ to include also a ratio of cumulants of the initial state eccentricity. However, this inevitably introduces uncertainties from modeling the initial state and it is therefore much preferable to compute $\W$ for collisions of the same nuclei at two different energies.

We performed further tests on how the observable performs in the presence of nonconformal effects.  Our nonconformal simulation results indicate that flow at LHC can be well estimated via a rescaling of conformal results by a constant factor of $0.8$. However, flow at RHIC exhibits a small additional centrality dependence of the scaling factor. Nevertheless, we found that by restricting the analysis to mid-central collisions, the same $\W$ observable can still be used to infer direct information about the interaction rate. Moreover, the interpretation of the results can be improved by supplying an appropriately adjusted theory calibration curve.

We also performed a first experimental extraction of the $\W$ observable by computing it from LHC data in PbPb collisions at $2.76\,$TeV and $5.02\,$TeV. Our results for $\W$ are in compatibility with the theory calibration curve, if the opacity is taken to be that of the \trento ~initial state model, and place PbPb collisions at LHC in a regime of opacities where hydrodynamics provides an accurate description of the space-time dynamics of the quark-gluon plasma.
This first application indicates that the observable $\W$ can indeed work in practice as intended.

Based on our analysis, we thus hope that 
future experimental determination of this observable can shed further light on the accuracy and limitations of hydrodynamic modeling of heavy-ion collisions. Clearly, this is particularly interesting when applied to OO collisions at RHIC and LHC, which, with the planned LHC run in 2025, will be the first time that the same nuclei were collided at the two facilities.

\begin{acknowledgments}
\textit{Acknowledgements:} 
This work is supported by the Deutsche Forschungsgemeinschaft (DFG, German Research Foundation)
through the CRC-TR 211 ’Strong-interaction matter under extreme conditions’– project
number 315477589 – TRR 211.
VV.E.A. gratefully acknowledges support by the European Union - NextGenerationEU through grant No. 760079/23.05.2023, funded by the Romanian ministry of research, innovation and digitization through Romania’s National Recovery and Resilience Plan, call no.~PNRR-III-C9-2022-I8.
C.W. was supported by the program Excellence Initiative–Research University of the University of Wrocław of the Ministry of Education and Science. C.W. has also received funding from the European Re-
search Council (ERC) under the European Union’s Horizon 2020 research and innovation programme (grant number: 101089093 / project acronym: High-TheQ). Views
and opinions expressed are however those of the authors
only and do not necessarily reflect those of the European
Union or the European Research Council. Neither the
European Union nor the granting authority can be held
responsible for them.
Numerical calculations presented in this work were performed at the Paderborn Center for Parallel Computing (PC2), the Center for Scientific Computing (CSC) at the Goethe-University of Frankfurt and 
the National Energy Research Scientific Computing Center (NERSC), a Department of Energy Office of Science User Facility, under the group name m2443 and we gratefully acknowledge their support. 
\end{acknowledgments}

\newpage

\clearpage

\newpage

\appendix

\section{Work function and improved calibration curve}\label{app:fwork_corr}

In order to discuss how the difference between initial and final state transverse energies affects theory predictions for the $\W$ observable (c.f. Eq.~\eqref{eq:W_def}), we first introduce another universal curve describing the system's dynamics. The opacity dependent work function is defined via the relative loss of transverse energy due to work performed in the longitudinal expansion over the whole evolution, i.e. it is given as
\begin{align}
    f_{\rm work}(\hat{\gamma})=\left.\frac{\d E_\perp}{\d \eta}\right/\frac{\d E^{(0)}_\perp}{\d \eta}\;.
\end{align}
Using this, we can obtain a more accurate estimate of the ratio of final state energies between two ensembles A and B with similar geometry, improving on Eq.~\eqref{eq:Eratio_estimate}.
\begin{align}
    \left(\left.\left\langle{\frac{\d E_\perp}{\d \eta}}\right\rangle_A\right/\left\langle{\frac{\d E_\perp}{\d \eta}}\right\rangle_B\right)^{1/4}=\frac{\langle{\hat{\gamma}}\rangle_Af_{\rm work}^{1/4}(\langle{\hat{\gamma}}\rangle_A)}{\langle{\hat{\gamma}}\rangle_Bf_{\rm work}^{1/4}(\langle{\hat{\gamma}}\rangle_B)}
\end{align}
Applying this to the definition of $\W$, we can find a better theory equivalent to compare to.
\begin{align}
    \W&=2 \cdot \frac{\log\left(c^E_{2}\{2\}|_A/c^E_{2}\{2\}|_B\right)}{\log\left(\left.\left\langle{\frac{\d E_\perp}{\d \eta}}\right\rangle_A\right/\left\langle{\frac{\d E_\perp}{\d \eta}}\right\rangle_B\right)} \\
    &\approx \frac{\Delta\log(\langle{\kappa}\rangle)}{\Delta\log(\langle{\hat{\gamma}}\rangle)+\frac{1}{4}\Delta\log(f_{\rm work}(\langle{\hat{\gamma}}\rangle))}\\
    &\approx \left.\frac{\d\log(\kappa)}{\d\log(\hat{\gamma})}\right/\left(1+\frac{1}{4}\frac{\d\log (f_{\rm work})}{\d\log(\hat{\gamma})}\right)\label{eq:calibration_improved}
\end{align}
The new expression now also involves the logarithmic derivative of the logarithmic work function. The asymptotic power law behaviour is known~\cite{Ambrus:2022koq}: ${\d\log (f_{\rm work})}/{\d\log(\hat{\gamma})}$ monotonically transitions from a value of $0$ in the free-streaming case to $-\frac{4}{9}$ for ideal hydrodynamics. This means that the term in the numerator does not have an effect on the asymptotic values, so they are the same as for ${\d\log (\kappa)}/{\d\log(\hat{\gamma})}$: 1 in the free-streaming case and 0 for ideal hydro. The term merely increases values in the transition regime by at most $11\%$.

\begin{figure}
    \centering
    \includegraphics[width=0.99\linewidth]{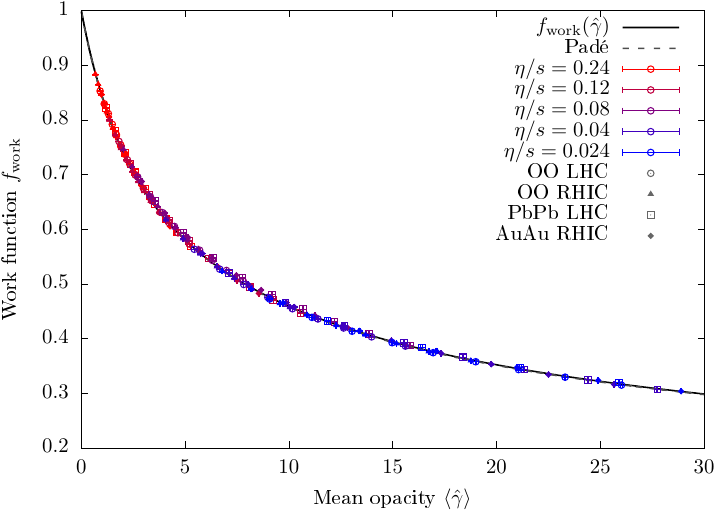}
    \caption{Work function $f_{\rm work}$, given as the mean value of the ratio $\left.\frac{\d E_\perp}{\d \eta}\right/\frac{\d E_\perp^{(0)}}{\d \eta}$ of final state to initial state transverse energies as a function of the mean opacity from our kinetic theory simulation results for centrality classes of size 10\% of OO collisions at LHC (open circles) and RHIC (filled triangles) as well as PbPb collisions (open squares) and AuAu collisions (filled diamonds) for various values of $\eta/s$ in color gradient from highest (red) to lowest (blue), compared to the work function $f_{\rm work}(\hat{\gamma})$ from our previous study (black) as well as a  Padé fit to the data (dark gray, dashed). Error bars show the standard error of the mean.}
    \label{fig:fwork}
\end{figure}

In order to obtain a calibration curve with the improved approximation, we now also need to extract the work function $f_{\rm work}(\hat{\gamma})$ from our results in a very similar way to how we extracted $\kappa(\hat{\gamma})$. Figure~\ref{fig:fwork} shows results for the mean value of $\left.\frac{\d E_\perp}{\d \eta}\right/\frac{\d E_\perp^{(0)}}{\d \eta}$ plotted against the mean opacity for centrality classes of size 10\%. Results for different collision systems are plotted with different symbols and different colors indicate different values of the shear viscosity. The black line shows an interpolation of $f_{\rm work}(\hat{\gamma})$ obtained from the results of our previous study~\cite{Ambrus:2022qya} with a fixed average initial state and the grey dashed line shows a Padé fit to the data points.

Since in this case the function that is to be approximated has a noninteger power in its limiting behaviour, the ansatz for the Padé fit has to be modified. The functional form that we decided on is 
\begin{align}
    f_{\rm work}^{\rm Pad\acute{e}}(\hat{\gamma})=\frac{1+(k_{\rm w,LO}+a)\hat{\gamma}+ck_{\rm w,id}\hat{\gamma}^2}{1+a\hat{\gamma}+b\hat{\gamma}^2+c\hat{\gamma}^{22/9}}\;.
\end{align}
The limiting cases of opacity-linear behavior and ideal hydrodynamics fix the two parameters controlling the asymptotic behaviour. In particular, this means $k_{\rm w,LO}=-0.212$ and $k_{\rm w,id}=1.64$. The remaining three parameters were determined via fitting to the plotted data to be $a=0.341(7)$, $b=0.0111(13)$ and $c=0.00477(90)$.

The data points for the mean ratio $\left.\frac{\d E_\perp}{\d \eta}\right/\frac{\d E_\perp^{(0)}}{\d \eta}$ line up almost perfectly. In conformal RTA kinetic theory, the work function seems to be universal to a very high accuracy, so long as means are taken over appropriate ensembles. Both the Padé fit and the work function from our previous study describe the data well. In particular, in the plot almost no difference between the two curves is visible. We have employed the Padé fit to compute the improved calibration curve that is displayed in all plots of results for $\W$.

\section{Flow response curve from hydrodynamic results}\label{app:hydro_curves}

\begin{figure}
    \centering
    \includegraphics[width=0.99\linewidth]{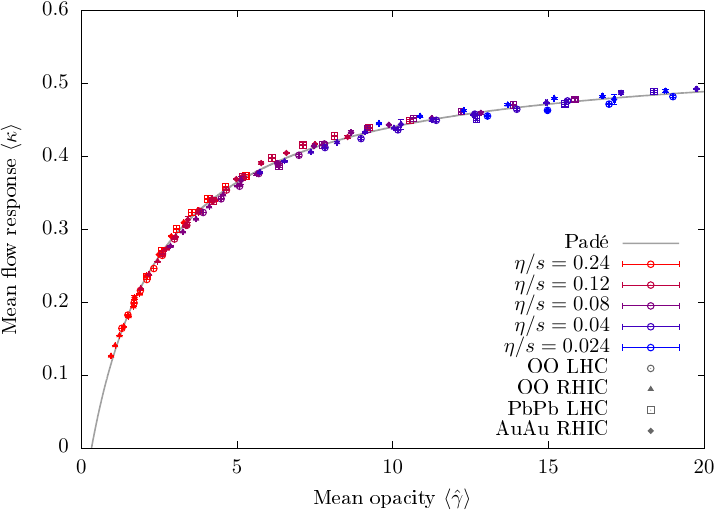}
    \caption{Mean values of the response $\kappa=\varepsilon_p/\epsilon_2$ as a function of the mean opacity from our hydrodynamic simulation results for centrality classes of size 10\% of OO collisions at LHC (open circles) and RHIC (filled triangles) as well as PbPb collisions (open squares) and AuAu collisions (filled diamonds) for various values of $\eta/s$ in color gradient from highest (red) to lowest (blue), compared to a Padé fit to the data (gray, solid). Error bars show the standard error of the mean.}
    \label{fig:kappa_hydro}
\end{figure}

In Sec.~\ref{sec:simulation_results} of this paper, we discussed the proposed $\W$ observable in terms of the flow response curve $\kappa(\hat{\gamma})$ that was obtained from simulation results in kinetic theory. The same curve can also be defined for hydrodynamic results, though the behavior is slightly different, as we have seen in our previous work~\cite{Ambrus:2022koq} and the companion paper~\cite{Ambrus:2024hks}. While the two curves agree at large opacities $\hat{\gamma}$, they start to diverge noticeably for $\hat{\gamma}\lesssim 4$, with hydrodynamic results for elliptic flow undershooting those of kinetic theory. At very low opacities $\hat{\gamma}\lesssim 0.5$, they even flip sign. The interpretation of this negative sign is that the flow anisotropy will develop its maximum in the same direction that the initial position space distribution had its maximum. This means that if we want to extract a Padé fit of the flow response curve from hydrodynamic data, we should adjust the ansatz of the fitting function. We choose it to be of the following form.

\begin{align}
    \kappa_{\rm hyd}(\hat{\gamma})=\frac{\kappa_{0,\rm hyd}+\kappa_{\rm LO,hyd}\hat{\gamma}}{1+\frac{\kappa_{\rm LO,hyd}}{\kappa_{\rm id}}\hat{\gamma}}
\end{align}

The ideal hydrodynamic response coefficient is the same as before, $\kappa_{\rm id}=0.547$. The other two coefficients were determined by a fit to the mean values from centrality classes of our hydrodynamic simulations to be $\kappa_{0,\rm hyd}=-0.0896(4)$ and $\kappa_{\rm LO,hyd}=0.271(3)$.

Figure~\ref{fig:kappa_hydro} shows our hydrodynamic simulation data for the mean values of $\varepsilon_p/\epsilon_2$ as a function of the mean opacity for our centrality classes in different collision systems plotted as different symbols and different values of $\eta/s$ plotted in different colors. The results clearly line up along a common curve, as in the case of kinetic theory. However, as anticipated, it has a slightly different form. While no simulation results in this work actually produced negative elliptic flow, the low opacity trend was captured well by our fitting curve allowing negative values.

\begin{figure}
    \centering
    \includegraphics[width=0.99\linewidth]{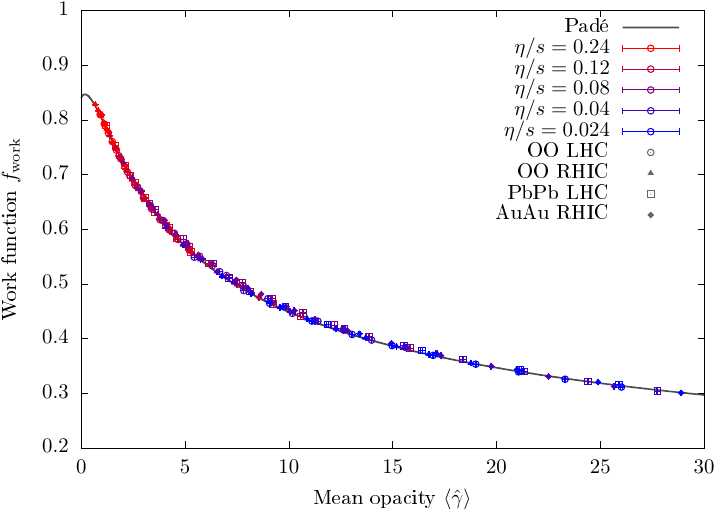}
    \caption{Work function $f_{\rm work}$, given as the mean values of the ratio $\left.\frac{\d E_\perp}{\d \eta}\right/\frac{\d E_\perp^{(0)}}{\d \eta}$ of final state to initial state transverse energies as a function of the mean opacity from our hydrodynamic simulation results for centrality classes of size 10\% of OO collisions at LHC (open circles) and RHIC (filled triangles) as well as PbPb collisions (open squares) and AuAu collisions (filled diamonds) for various values of $\eta/s$ in color gradient from highest (red) to lowest (blue), compared to the work function $f_{\rm work}(\hat{\gamma})$ from our previous study (black) as well as a  Padé fit to the data (dark gray, dashed). Error bars show the standard error of the mean.}
    \label{fig:fwork_hydro}
\end{figure}

Next, we can also consider the $f_{\rm work}(\hat{\gamma})$-curve in hydrodynamics. It should again be mostly similar but deviate slightly at small opacities. In particular, our initialization scheme of hydrodynamics reduces the initial transverse energy $\epni$ in anticipation of a rise in Bjorken-flow--like pre-equilibrium~\cite{Ambrus:2022koq}, but for small opacities the system never equilibrates fully, such that $\d E_\perp/\d\eta$ remains small. Our data in this work do not reach to low enough opacities to fully capture the behavior, so we choose a very simple ansatz for the Padé fit that is not necessarily accurate at lower opacities.
\begin{align}
    f_{\rm work}^{\rm hyd}(\hat{\gamma})=\frac{f_{0,\rm hyd}+c_{\rm hyd}k_{\rm w,id}\hat{\gamma}}{1+a_{\rm hyd}\hat{\gamma}+c_{\rm hyd}\hat{\gamma}^{13/9}}
\end{align}

Again, the ideal hydrodynamic limit is fixed as before, $k_{\rm w,id}=1.64$. For the other coefficients, we obtained values of $f_{0,\rm hyd}=0.840(2)$, $a_{\rm hyd}=0.280(7)$ and $c_{\rm hyd}=0.221(8)$.

In Figure~\ref{fig:fwork_hydro}, we present our hydrodynamic results for the mean values of $\left.\frac{\d E_\perp}{\d\eta}\right/\epn$ as a function of the mean opacity for centrality classes of different collision systems as different symbols and different values of $\eta/s$ in different colors. While the Padé curve is likely inaccurate at opacities $\hat{\gamma}\lesssim 1$, for which we do not have simulation results in this work, the data that we do have is described well, so it can serve as an approximate parametrization of the work function $f_{\rm work}(\hat{\gamma})$.

\begin{figure}
    \centering
    \includegraphics[width=0.99\linewidth]{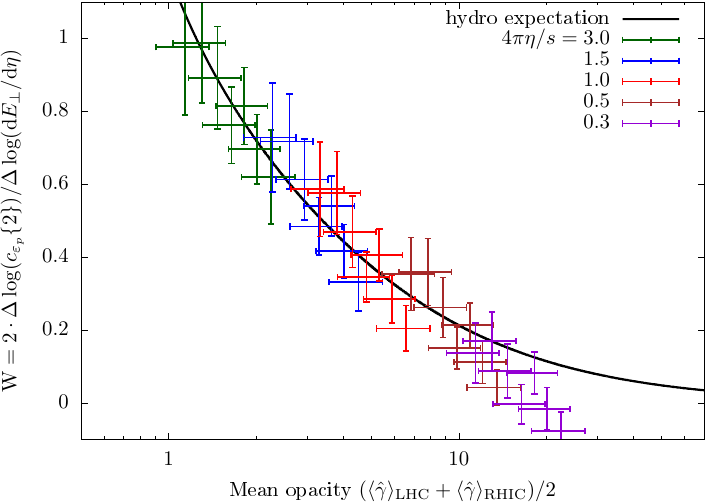}
    \caption{$\W$ computed from hydrodynamic simulation results of centrality classes of size 10\% up to the 60-70\% class for the OO collision systems at RHIC and LHC at different shear viscosities $\eta/s$ in different colors, compared to the theory expectation curve (black) extracted from the Padé fits to hydrodynamic data of $\kappa(\hat{\gamma})$ and $f_{\rm work}(\hat{\gamma})$. Error bars show the difference in opacity between the two systems and the propagated error of the cumulants and transverse energies.}
    \label{fig:Whyd}
\end{figure}

Combining the two curves, it is technically also possible to extract a calibration curve for our hydrodynamization observable $\W$. However, the negative and, in particular, nonvanishing elliptic flow at small opacities contradicts the expectation that in this regime flow should build up linearly in the opacity. Thus, instead of a low opacity limit of $\lim_{\hat{\gamma}\to 0}{\d\log(\kappa)}/{\d\log(\hat{\gamma})}=1$, in hydrodynamics, this quantity diverges when approaching the zero crossing of $\kappa(\hat{\gamma})$ and becomes undefined at even lower opacities. We stress that at these low opacities the system is too dilute for hydrodynamics to be accurate and that in reality the linear order limit of $\W$ should be expected. Nevertheless, we can still examine how the hydrodynamic results behave.

In Figure~\ref{fig:Whyd}, we perform a comparison of the theory expectation curve and hydrodynamic data from simulations of OO collisions at RHIC and LHC. The points show values of $\W$ as computed from the RHIC and LHC results from the same centrality class, for classes of size 10\% up to the 60-70\% centrality class. Different colors indicate different values of $\eta/s$. The black line is the theory expectation as computed from the two Padé fits according to Eq.~\eqref{eq:calibration_improved}. While results at low opacities show unreasonably large values of $\W$, they fall in line exactly with the expectation. This means that the computation of $\W$ from hydrodynamic results goes wrong in a predictable way and is due to the lower value of the flow response coefficient compared to kinetic theory at the same opacity.

\bibliography{main.bib}

\begin{thebibliography}{25}%
\makeatletter
\providecommand \@ifxundefined [1]{%
 \@ifx{#1\undefined}
}%
\providecommand \@ifnum [1]{%
 \ifnum #1\expandafter \@firstoftwo
 \else \expandafter \@secondoftwo
 \fi
}%
\providecommand \@ifx [1]{%
 \ifx #1\expandafter \@firstoftwo
 \else \expandafter \@secondoftwo
 \fi
}%
\providecommand \natexlab [1]{#1}%
\providecommand \enquote  [1]{``#1''}%
\providecommand \bibnamefont  [1]{#1}%
\providecommand \bibfnamefont [1]{#1}%
\providecommand \citenamefont [1]{#1}%
\providecommand \href@noop [0]{\@secondoftwo}%
\providecommand \href [0]{\begingroup \@sanitize@url \@href}%
\providecommand \@href[1]{\@@startlink{#1}\@@href}%
\providecommand \@@href[1]{\endgroup#1\@@endlink}%
\providecommand \@sanitize@url [0]{\catcode `\\12\catcode `\$12\catcode `\&12\catcode `\#12\catcode `\^12\catcode `\_12\catcode `\%12\relax}%
\providecommand \@@startlink[1]{}%
\providecommand \@@endlink[0]{}%
\providecommand \url  [0]{\begingroup\@sanitize@url \@url }%
\providecommand \@url [1]{\endgroup\@href {#1}{\urlprefix }}%
\providecommand \urlprefix  [0]{URL }%
\providecommand \Eprint [0]{\href }%
\providecommand \doibase [0]{http://dx.doi.org/}%
\providecommand \selectlanguage [0]{\@gobble}%
\providecommand \bibinfo  [0]{\@secondoftwo}%
\providecommand \bibfield  [0]{\@secondoftwo}%
\providecommand \translation [1]{[#1]}%
\providecommand \BibitemOpen [0]{}%
\providecommand \bibitemStop [0]{}%
\providecommand \bibitemNoStop [0]{.\EOS\space}%
\providecommand \EOS [0]{\spacefactor3000\relax}%
\providecommand \BibitemShut  [1]{\csname bibitem#1\endcsname}%
\let\auto@bib@innerbib\@empty
\bibitem [{\citenamefont {Abelev}\ \emph {et~al.}(2014)\citenamefont {Abelev} \emph {et~al.}}]{ALICE:2014dwt}%
  \BibitemOpen
  \bibfield  {author} {\bibinfo {author} {\bibfnamefont {B.~B.}\ \bibnamefont {Abelev}} \emph {et~al.} (\bibinfo {collaboration} {ALICE}),\ }\href {\doibase 10.1103/PhysRevC.90.054901} {\bibfield  {journal} {\bibinfo  {journal} {Phys. Rev. C}\ }\textbf {\bibinfo {volume} {90}},\ \bibinfo {pages} {054901} (\bibinfo {year} {2014})},\ \Eprint {http://arxiv.org/abs/1406.2474} {arXiv:1406.2474 [nucl-ex]} \BibitemShut {NoStop}%
\bibitem [{\citenamefont {Aaboud}\ \emph {et~al.}(2017)\citenamefont {Aaboud} \emph {et~al.}}]{ATLAS:2017hap}%
  \BibitemOpen
  \bibfield  {author} {\bibinfo {author} {\bibfnamefont {M.}~\bibnamefont {Aaboud}} \emph {et~al.} (\bibinfo {collaboration} {ATLAS}),\ }\href {\doibase 10.1140/epjc/s10052-017-4988-1} {\bibfield  {journal} {\bibinfo  {journal} {Eur. Phys. J. C}\ }\textbf {\bibinfo {volume} {77}},\ \bibinfo {pages} {428} (\bibinfo {year} {2017})},\ \Eprint {http://arxiv.org/abs/1705.04176} {arXiv:1705.04176 [hep-ex]} \BibitemShut {NoStop}%
\bibitem [{\citenamefont {Sirunyan}\ \emph {et~al.}(2018)\citenamefont {Sirunyan} \emph {et~al.}}]{CMS:2017kcs}%
  \BibitemOpen
  \bibfield  {author} {\bibinfo {author} {\bibfnamefont {A.~M.}\ \bibnamefont {Sirunyan}} \emph {et~al.} (\bibinfo {collaboration} {CMS}),\ }\href {\doibase 10.1103/PhysRevLett.120.092301} {\bibfield  {journal} {\bibinfo  {journal} {Phys. Rev. Lett.}\ }\textbf {\bibinfo {volume} {120}},\ \bibinfo {pages} {092301} (\bibinfo {year} {2018})},\ \Eprint {http://arxiv.org/abs/1709.09189} {arXiv:1709.09189 [nucl-ex]} \BibitemShut {NoStop}%
\bibitem [{\citenamefont {Grosse-Oetringhaus}\ and\ \citenamefont {Wiedemann}(2024)}]{Grosse-Oetringhaus:2024bwr}%
  \BibitemOpen
  \bibfield  {author} {\bibinfo {author} {\bibfnamefont {J.~F.}\ \bibnamefont {Grosse-Oetringhaus}}\ and\ \bibinfo {author} {\bibfnamefont {U.~A.}\ \bibnamefont {Wiedemann}},\ }\href@noop {} {\  (\bibinfo {year} {2024})},\ \Eprint {http://arxiv.org/abs/2407.07484} {arXiv:2407.07484 [hep-ex]} \BibitemShut {NoStop}%
\bibitem [{\citenamefont {Ambrus}\ \emph {et~al.}(2023{\natexlab{a}})\citenamefont {Ambrus}, \citenamefont {Schlichting},\ and\ \citenamefont {Werthmann}}]{Ambrus:2022koq}%
  \BibitemOpen
  \bibfield  {author} {\bibinfo {author} {\bibfnamefont {V.~E.}\ \bibnamefont {Ambrus}}, \bibinfo {author} {\bibfnamefont {S.}~\bibnamefont {Schlichting}}, \ and\ \bibinfo {author} {\bibfnamefont {C.}~\bibnamefont {Werthmann}},\ }\href {\doibase 10.1103/PhysRevD.107.094013} {\bibfield  {journal} {\bibinfo  {journal} {Phys. Rev. D}\ }\textbf {\bibinfo {volume} {107}},\ \bibinfo {pages} {094013} (\bibinfo {year} {2023}{\natexlab{a}})},\ \Eprint {http://arxiv.org/abs/2211.14379} {arXiv:2211.14379 [hep-ph]} \BibitemShut {NoStop}%
\bibitem [{\citenamefont {Ambrus}\ \emph {et~al.}(2023{\natexlab{b}})\citenamefont {Ambrus}, \citenamefont {Schlichting},\ and\ \citenamefont {Werthmann}}]{Ambrus:2022qya}%
  \BibitemOpen
  \bibfield  {author} {\bibinfo {author} {\bibfnamefont {V.~E.}\ \bibnamefont {Ambrus}}, \bibinfo {author} {\bibfnamefont {S.}~\bibnamefont {Schlichting}}, \ and\ \bibinfo {author} {\bibfnamefont {C.}~\bibnamefont {Werthmann}},\ }\href {\doibase 10.1103/PhysRevLett.130.152301} {\bibfield  {journal} {\bibinfo  {journal} {Phys. Rev. Lett.}\ }\textbf {\bibinfo {volume} {130}},\ \bibinfo {pages} {152301} (\bibinfo {year} {2023}{\natexlab{b}})},\ \Eprint {http://arxiv.org/abs/2211.14356} {arXiv:2211.14356 [hep-ph]} \BibitemShut {NoStop}%
\bibitem [{\citenamefont {Ambrus}\ \emph {et~al.}(2024)\citenamefont {Ambrus}, \citenamefont {Schlichting},\ and\ \citenamefont {Werthmann}}]{Ambrus:2024hks}%
  \BibitemOpen
  \bibfield  {author} {\bibinfo {author} {\bibfnamefont {V.~E.}\ \bibnamefont {Ambrus}}, \bibinfo {author} {\bibfnamefont {S.}~\bibnamefont {Schlichting}}, \ and\ \bibinfo {author} {\bibfnamefont {C.}~\bibnamefont {Werthmann}},\ }\href@noop {} {\  (\bibinfo {year} {2024})},\ \Eprint {http://arxiv.org/abs/2411.19708} {arXiv:2411.19708 [hep-ph]} \BibitemShut {NoStop}%
\bibitem [{\citenamefont {Werthmann}\ \emph {et~al.}(2025)\citenamefont {Werthmann}, \citenamefont {Ambrus},\ and\ \citenamefont {Schlichting}}]{werthmann_2025_14849764}%
  \BibitemOpen
  \bibfield  {author} {\bibinfo {author} {\bibfnamefont {C.}~\bibnamefont {Werthmann}}, \bibinfo {author} {\bibfnamefont {V.~E.}\ \bibnamefont {Ambrus}}, \ and\ \bibinfo {author} {\bibfnamefont {S.}~\bibnamefont {Schlichting}},\ }\href {\doibase 10.5281/zenodo.14849764} {\enquote {\bibinfo {title} {Plot data for "collective dynamics in heavy and light-ion collisions -- ii) determining the origin of collective behavior in high-energy collisions"},}\ } (\bibinfo {year} {2025})\BibitemShut {NoStop}%
\bibitem [{\citenamefont {Lim}\ \emph {et~al.}(2019)\citenamefont {Lim}, \citenamefont {Carlson}, \citenamefont {Loizides}, \citenamefont {Lonardoni}, \citenamefont {Lynn}, \citenamefont {Nagle}, \citenamefont {Orjuela~Koop},\ and\ \citenamefont {Ouellette}}]{Lim:2018huo}%
  \BibitemOpen
  \bibfield  {author} {\bibinfo {author} {\bibfnamefont {S.~H.}\ \bibnamefont {Lim}}, \bibinfo {author} {\bibfnamefont {J.}~\bibnamefont {Carlson}}, \bibinfo {author} {\bibfnamefont {C.}~\bibnamefont {Loizides}}, \bibinfo {author} {\bibfnamefont {D.}~\bibnamefont {Lonardoni}}, \bibinfo {author} {\bibfnamefont {J.~E.}\ \bibnamefont {Lynn}}, \bibinfo {author} {\bibfnamefont {J.~L.}\ \bibnamefont {Nagle}}, \bibinfo {author} {\bibfnamefont {J.~D.}\ \bibnamefont {Orjuela~Koop}}, \ and\ \bibinfo {author} {\bibfnamefont {J.}~\bibnamefont {Ouellette}},\ }\href {\doibase 10.1103/PhysRevC.99.044904} {\bibfield  {journal} {\bibinfo  {journal} {Phys. Rev. C}\ }\textbf {\bibinfo {volume} {99}},\ \bibinfo {pages} {044904} (\bibinfo {year} {2019})},\ \Eprint {http://arxiv.org/abs/1812.08096} {arXiv:1812.08096 [nucl-th]} \BibitemShut {NoStop}%
\bibitem [{TGl()}]{TGlauberMC}%
  \BibitemOpen
  \href@noop {} {\enquote {\bibinfo {title} {{TGlauberMC on HepForge}.}}\ }\bibinfo {howpublished} {\url{https://tglaubermc.hepforge.org}},\ \bibinfo {note} {v3.2}\BibitemShut {NoStop}%
\bibitem [{\citenamefont {Rybczy\'nski}\ and\ \citenamefont {Broniowski}(2019)}]{Rybczynski:2019adt}%
  \BibitemOpen
  \bibfield  {author} {\bibinfo {author} {\bibfnamefont {M.}~\bibnamefont {Rybczy\'nski}}\ and\ \bibinfo {author} {\bibfnamefont {W.}~\bibnamefont {Broniowski}},\ }\href {\doibase 10.1103/PhysRevC.100.064912} {\bibfield  {journal} {\bibinfo  {journal} {Phys. Rev. C}\ }\textbf {\bibinfo {volume} {100}},\ \bibinfo {pages} {064912} (\bibinfo {year} {2019})},\ \Eprint {http://arxiv.org/abs/1910.09489} {arXiv:1910.09489 [hep-ph]} \BibitemShut {NoStop}%
\bibitem [{\citenamefont {Nijs}\ and\ \citenamefont {van~der Schee}(2022)}]{Nijs:2021clz}%
  \BibitemOpen
  \bibfield  {author} {\bibinfo {author} {\bibfnamefont {G.}~\bibnamefont {Nijs}}\ and\ \bibinfo {author} {\bibfnamefont {W.}~\bibnamefont {van~der Schee}},\ }\href {\doibase 10.1103/PhysRevC.106.044903} {\bibfield  {journal} {\bibinfo  {journal} {Phys. Rev. C}\ }\textbf {\bibinfo {volume} {106}},\ \bibinfo {pages} {044903} (\bibinfo {year} {2022})},\ \Eprint {http://arxiv.org/abs/2110.13153} {arXiv:2110.13153 [nucl-th]} \BibitemShut {NoStop}%
\bibitem [{\citenamefont {Ambru\cb{s}}\ \emph {et~al.}(2022)\citenamefont {Ambru\cb{s}}, \citenamefont {Schlichting},\ and\ \citenamefont {Werthmann}}]{Ambrus:2021fej}%
  \BibitemOpen
  \bibfield  {author} {\bibinfo {author} {\bibfnamefont {V.~E.}\ \bibnamefont {Ambru\cb{s}}}, \bibinfo {author} {\bibfnamefont {S.}~\bibnamefont {Schlichting}}, \ and\ \bibinfo {author} {\bibfnamefont {C.}~\bibnamefont {Werthmann}},\ }\href {\doibase 10.1103/PhysRevD.105.014031} {\bibfield  {journal} {\bibinfo  {journal} {Phys. Rev. D}\ }\textbf {\bibinfo {volume} {105}},\ \bibinfo {pages} {014031} (\bibinfo {year} {2022})},\ \Eprint {http://arxiv.org/abs/2109.03290} {arXiv:2109.03290 [hep-ph]} \BibitemShut {NoStop}%
\bibitem [{\citenamefont {Bhalerao}\ \emph {et~al.}(2011)\citenamefont {Bhalerao}, \citenamefont {Luzum},\ and\ \citenamefont {Ollitrault}}]{Bhalerao:2011yg}%
  \BibitemOpen
  \bibfield  {author} {\bibinfo {author} {\bibfnamefont {R.~S.}\ \bibnamefont {Bhalerao}}, \bibinfo {author} {\bibfnamefont {M.}~\bibnamefont {Luzum}}, \ and\ \bibinfo {author} {\bibfnamefont {J.-Y.}\ \bibnamefont {Ollitrault}},\ }\href {\doibase 10.1103/PhysRevC.84.034910} {\bibfield  {journal} {\bibinfo  {journal} {Phys. Rev. C}\ }\textbf {\bibinfo {volume} {84}},\ \bibinfo {pages} {034910} (\bibinfo {year} {2011})},\ \Eprint {http://arxiv.org/abs/1104.4740} {arXiv:1104.4740 [nucl-th]} \BibitemShut {NoStop}%
\bibitem [{\citenamefont {Giacalone}\ \emph {et~al.}(2017)\citenamefont {Giacalone}, \citenamefont {Yan}, \citenamefont {Noronha-Hostler},\ and\ \citenamefont {Ollitrault}}]{Giacalone:2016eyu}%
  \BibitemOpen
  \bibfield  {author} {\bibinfo {author} {\bibfnamefont {G.}~\bibnamefont {Giacalone}}, \bibinfo {author} {\bibfnamefont {L.}~\bibnamefont {Yan}}, \bibinfo {author} {\bibfnamefont {J.}~\bibnamefont {Noronha-Hostler}}, \ and\ \bibinfo {author} {\bibfnamefont {J.-Y.}\ \bibnamefont {Ollitrault}},\ }\href {\doibase 10.1103/PhysRevC.95.014913} {\bibfield  {journal} {\bibinfo  {journal} {Phys. Rev. C}\ }\textbf {\bibinfo {volume} {95}},\ \bibinfo {pages} {014913} (\bibinfo {year} {2017})},\ \Eprint {http://arxiv.org/abs/1608.01823} {arXiv:1608.01823 [nucl-th]} \BibitemShut {NoStop}%
\bibitem [{\citenamefont {Acharya}\ \emph {et~al.}(2018)\citenamefont {Acharya} \emph {et~al.}}]{ALICE:2018rtz}%
  \BibitemOpen
  \bibfield  {author} {\bibinfo {author} {\bibfnamefont {S.}~\bibnamefont {Acharya}} \emph {et~al.} (\bibinfo {collaboration} {ALICE}),\ }\href {\doibase 10.1007/JHEP07(2018)103} {\bibfield  {journal} {\bibinfo  {journal} {JHEP}\ }\textbf {\bibinfo {volume} {07}},\ \bibinfo {pages} {103} (\bibinfo {year} {2018})},\ \Eprint {http://arxiv.org/abs/1804.02944} {arXiv:1804.02944 [nucl-ex]} \BibitemShut {NoStop}%
\bibitem [{\citenamefont {Acharya}\ \emph {et~al.}(2023)\citenamefont {Acharya} \emph {et~al.}}]{ALICE:2022imr}%
  \BibitemOpen
  \bibfield  {author} {\bibinfo {author} {\bibfnamefont {S.}~\bibnamefont {Acharya}} \emph {et~al.} (\bibinfo {collaboration} {ALICE}),\ }\href {\doibase 10.1016/j.physletb.2023.137730} {\bibfield  {journal} {\bibinfo  {journal} {Phys. Lett. B}\ }\textbf {\bibinfo {volume} {845}},\ \bibinfo {pages} {137730} (\bibinfo {year} {2023})},\ \Eprint {http://arxiv.org/abs/2204.10210} {arXiv:2204.10210 [nucl-ex]} \BibitemShut {NoStop}%
\bibitem [{\citenamefont {Singh}\ \emph {et~al.}(2024)\citenamefont {Singh}, \citenamefont {Giacalone}, \citenamefont {Schenke},\ and\ \citenamefont {Schlichting}}]{Singh:2023rkg}%
  \BibitemOpen
  \bibfield  {author} {\bibinfo {author} {\bibfnamefont {P.}~\bibnamefont {Singh}}, \bibinfo {author} {\bibfnamefont {G.}~\bibnamefont {Giacalone}}, \bibinfo {author} {\bibfnamefont {B.}~\bibnamefont {Schenke}}, \ and\ \bibinfo {author} {\bibfnamefont {S.}~\bibnamefont {Schlichting}},\ }\href {\doibase 10.1051/epjconf/202429610005} {\bibfield  {journal} {\bibinfo  {journal} {EPJ Web Conf.}\ }\textbf {\bibinfo {volume} {296}},\ \bibinfo {pages} {10005} (\bibinfo {year} {2024})},\ \Eprint {http://arxiv.org/abs/2312.07462} {arXiv:2312.07462 [hep-ph]} \BibitemShut {NoStop}%
\bibitem [{\citenamefont {M\"antysaari}\ and\ \citenamefont {Singh}(2024)}]{Mantysaari:2024qmt}%
  \BibitemOpen
  \bibfield  {author} {\bibinfo {author} {\bibfnamefont {H.}~\bibnamefont {M\"antysaari}}\ and\ \bibinfo {author} {\bibfnamefont {P.}~\bibnamefont {Singh}},\ }\href@noop {} {\  (\bibinfo {year} {2024})},\ \Eprint {http://arxiv.org/abs/2411.14934} {arXiv:2411.14934 [nucl-th]} \BibitemShut {NoStop}%
\bibitem [{\citenamefont {Kurkela}\ \emph {et~al.}(2019)\citenamefont {Kurkela}, \citenamefont {Wiedemann},\ and\ \citenamefont {Wu}}]{Kurkela:2019kip}%
  \BibitemOpen
  \bibfield  {author} {\bibinfo {author} {\bibfnamefont {A.}~\bibnamefont {Kurkela}}, \bibinfo {author} {\bibfnamefont {U.~A.}\ \bibnamefont {Wiedemann}}, \ and\ \bibinfo {author} {\bibfnamefont {B.}~\bibnamefont {Wu}},\ }\href {\doibase 10.1140/epjc/s10052-019-7428-6} {\bibfield  {journal} {\bibinfo  {journal} {Eur. Phys. J. C}\ }\textbf {\bibinfo {volume} {79}},\ \bibinfo {pages} {965} (\bibinfo {year} {2019})},\ \Eprint {http://arxiv.org/abs/1905.05139} {arXiv:1905.05139 [hep-ph]} \BibitemShut {NoStop}%
\bibitem [{\citenamefont {Karpenko}\ \emph {et~al.}(2014)\citenamefont {Karpenko}, \citenamefont {Huovinen},\ and\ \citenamefont {Bleicher}}]{Karpenko:2013wva}%
  \BibitemOpen
  \bibfield  {author} {\bibinfo {author} {\bibfnamefont {I.}~\bibnamefont {Karpenko}}, \bibinfo {author} {\bibfnamefont {P.}~\bibnamefont {Huovinen}}, \ and\ \bibinfo {author} {\bibfnamefont {M.}~\bibnamefont {Bleicher}},\ }\href {\doibase 10.1016/j.cpc.2014.07.010} {\bibfield  {journal} {\bibinfo  {journal} {Comput. Phys. Commun.}\ }\textbf {\bibinfo {volume} {185}},\ \bibinfo {pages} {3016} (\bibinfo {year} {2014})},\ \Eprint {http://arxiv.org/abs/1312.4160} {arXiv:1312.4160 [nucl-th]} \BibitemShut {NoStop}%
\bibitem [{\citenamefont {Steinheimer}\ \emph {et~al.}(2011)\citenamefont {Steinheimer}, \citenamefont {Schramm},\ and\ \citenamefont {Stocker}}]{Steinheimer:2010ib}%
  \BibitemOpen
  \bibfield  {author} {\bibinfo {author} {\bibfnamefont {J.}~\bibnamefont {Steinheimer}}, \bibinfo {author} {\bibfnamefont {S.}~\bibnamefont {Schramm}}, \ and\ \bibinfo {author} {\bibfnamefont {H.}~\bibnamefont {Stocker}},\ }\href {\doibase 10.1088/0954-3899/38/3/035001} {\bibfield  {journal} {\bibinfo  {journal} {J. Phys. G}\ }\textbf {\bibinfo {volume} {38}},\ \bibinfo {pages} {035001} (\bibinfo {year} {2011})},\ \Eprint {http://arxiv.org/abs/1009.5239} {arXiv:1009.5239 [hep-ph]} \BibitemShut {NoStop}%
\bibitem [{\citenamefont {Zhang}\ \emph {et~al.}(2024)\citenamefont {Zhang}, \citenamefont {Chen}, \citenamefont {Giacalone}, \citenamefont {Huang}, \citenamefont {Jia},\ and\ \citenamefont {Ma}}]{Zhang:2024vkh}%
  \BibitemOpen
  \bibfield  {author} {\bibinfo {author} {\bibfnamefont {C.}~\bibnamefont {Zhang}}, \bibinfo {author} {\bibfnamefont {J.}~\bibnamefont {Chen}}, \bibinfo {author} {\bibfnamefont {G.}~\bibnamefont {Giacalone}}, \bibinfo {author} {\bibfnamefont {S.}~\bibnamefont {Huang}}, \bibinfo {author} {\bibfnamefont {J.}~\bibnamefont {Jia}}, \ and\ \bibinfo {author} {\bibfnamefont {Y.-G.}\ \bibnamefont {Ma}},\ }\href@noop {} {\  (\bibinfo {year} {2024})},\ \Eprint {http://arxiv.org/abs/2404.08385} {arXiv:2404.08385 [nucl-th]} \BibitemShut {NoStop}%
\bibitem [{\citenamefont {Giacalone}\ \emph {et~al.}(2024{\natexlab{a}})\citenamefont {Giacalone} \emph {et~al.}}]{Giacalone:2024luz}%
  \BibitemOpen
  \bibfield  {author} {\bibinfo {author} {\bibfnamefont {G.}~\bibnamefont {Giacalone}} \emph {et~al.},\ }\href@noop {} {\  (\bibinfo {year} {2024}{\natexlab{a}})},\ \Eprint {http://arxiv.org/abs/2402.05995} {arXiv:2402.05995 [nucl-th]} \BibitemShut {NoStop}%
\bibitem [{\citenamefont {Giacalone}\ \emph {et~al.}(2024{\natexlab{b}})\citenamefont {Giacalone} \emph {et~al.}}]{Giacalone:2024ixe}%
  \BibitemOpen
  \bibfield  {author} {\bibinfo {author} {\bibfnamefont {G.}~\bibnamefont {Giacalone}} \emph {et~al.},\ }\href@noop {} {\  (\bibinfo {year} {2024}{\natexlab{b}})},\ \Eprint {http://arxiv.org/abs/2405.20210} {arXiv:2405.20210 [nucl-th]} \BibitemShut {NoStop}%
\end{thebibliography}%

\end{document}